\documentclass[journal,draftcls,onecolumn,12pt,twoside]{IEEEtran}
\pdfoutput=1
\usepackage{cite}

\ifCLASSINFOpdf
   \usepackage[pdftex]{graphicx}

\else
  
\fi

\usepackage[cmex10]{amsmath}
\usepackage{amssymb} 
\usepackage{amsopn} 

\usepackage{array}

\usepackage{enumerate}

\newtheorem{theorem}{Theorem} 
\newtheorem{lemma}{Lemma}
\newtheorem{remark}{Remark}
\newtheorem{definition}{Definition}
\newtheorem{corollary}{Corollary}

\usepackage[usenames,dvipsnames]{xcolor}
\usepackage{hyperref}
\hypersetup{colorlinks=true,linkcolor=BlueViolet, citecolor=PineGreen,urlcolor=black}

\usepackage{xcolor}

\newcommand{\Wf}{0.9} 

\usepackage{caption}
\usepackage{subcaption}

\begin{document}
%

\title{Robust Successive Compute-and-Forward\\ over Multi-User Multi-Relay Networks}

\author{Mohsen~Hejazi, Seyed~Mohammad~Azimi-Abarghouyi, Behrooz~Makki, Masoumeh~Nasiri-Kenari,~\IEEEmembership{Senior~Member,~IEEE,} and~Tommy~Svensson,~\IEEEmembership{Senior~Member,~IEEE}
\thanks{M. Hejazi, S. M. Azimi-Abarghouyi, and M. Nasiri-Kenari are with the Wireless Research Laboratory (WRL), Electrical
 Engineering Department, Sharif University of Technology, Tehran,
  Iran (e-mail: \{mhejazi, sm\_azimi\}@ee.sharif.edu; mnasiri@sharif.edu).}
\thanks{B.~Makki and T.~Svensson are with the Department of Signals and Systems, Chalmers University of Technology, Gothenburg, Sweden (email: \{behrooz.makki, tommy.svensson\}@chalmers.se).}   
}



\maketitle

\begin{abstract}
This paper develops efficient Compute-and-forward (CMF) schemes in multi-user multi-relay networks.
To solve the rank failure problem in CMF setups and to achieve full diversity of the network, we introduce two novel CMF methods, namely, extended CMF and successive CMF. 
The former, having low complexity, is based on  recovering multiple equations at relays. The latter utilizes successive interference cancellation (SIC) to enhance the system performance compared to the state-of-the-art schemes.
Both methods can be utilized in a network with different number of users, relays, and relay antennas, 
with negligible feedback channels or signaling overhead.
We derive new concise formulations and explicit framework for the successive CMF method as well as an approach to reduce its computational complexity.
Our theoretical analysis and computer simulations demonstrate the superior performance of our proposed CMF methods over the conventional schemes. 
Furthermore, based on our simulation results, the successive CMF method yields additional signal-to-noise ratio gains and shows considerable robustness against channel estimation error, compared to the extended CMF method.
\end{abstract}

\begin{IEEEkeywords}
Multi-user relay network, Wireless relay network, Successive compute-and-forward, Outage probability, Performance analysis, Multi-antenna relay.
\end{IEEEkeywords}

\IEEEpeerreviewmaketitle

\section{Introduction}  \label{sec:intro}
\IEEEPARstart{T}{he} compute-and-forward (CMF) method, proposed by Nazer and Gastpar~\cite{BN1}, is an innovative approach for efficient communications over multi-user relay networks. Here, instead of recovering single messages, the relays attempt to reliably decode (compute) and pass an integer linear combination of the transmitted messages, referred to as an equation, to the destination. By receiving sufficient equations and their corresponding equation coefficients vectors (ECVs), the destination can solve the linear equation system to recover the desired messages. The CMF method enables exploiting, rather than combating, the multiple access interference in a wireless relay network, and thus results in improved network throughput~\cite{BN2}.

A ``CMF method'' designed for a multi-user multi-relay network consists of two main parts, namely, ``computing scheme'' and ``forwarding strategy''. Computing scheme is the structure that is used in each relay to find ECVs and to compute the desired equations from the received signals by the relays. On the other hand, forwarding strategy determines the plan of exchanging information over the network, e.g. how to transmit decoded equations by the relays to the destination, and how to recover the desired messages at the destination.

\begin{table*}[t]
\begin{minipage}{\textwidth}
\setlength{\tabcolsep}{3pt}
\centering
\caption{Comparison of the CMF methods.}
\begin{tabular}{p{13mm} c c ccccccccc p{30mm}}
\hline
\hline
\multicolumn{3}{c}{CMF method}& & \multicolumn{4}{c}{Performance\footnote{The Parameters $d_\text{Id}$ and $d_\text{WF}$ denote the diversity orders corresponding to the cases of ideal and with-feedback Nakagami($q$) R-D channels, respectively (see Theorems~\ref{th:d-Ideal} and~\ref{th:d-WF}). Also, based on Fig.~\ref{fig:CEE}, $\gamma_{_{\text{Id}}}$ and $\gamma_{_\text{CEE}}$ are the required SNRs for achieving outage probability  of 0.01 for the cases of CEE variances equal to 0 and 0.05, respectively.}} & &   \multicolumn{3}{c}{Complexity\footnote{\#TS: Number of time slots per transmission frame. \#ECV: Number of calculated ECVs in all relays. \#Eq: Number of computed equations in all relays.}}  &  \\

\cline{1-3} \cline{5-8} \cline{10-12}
Name & Computing & Forwarding & & $d_\text{Id}$ & $d_\text{WF}$ & $\gamma_{_{\text{Id}}}$ & $\gamma_{_\text{CEE}}$ & & \#TS & \#ECV & \#Eq & \multicolumn{1}{c}{Notes} \vspace{-1mm} \\

& scheme & strategy & & & & & & & & & & \\

\hline 
Original CMF~\cite{BN1} & Std-CM & Std-FW & & $\cong 0$ & $\cong 0$ & -- & -- & & $M+1$ & $M$ & $M$ & -Rank Failure Problem, \newline -Requires $M \geq L$, \newline -Sensetive to CEE. \\ 

\hline 
Extended CMF & Ext-CM & Sel-FW & & $\frac{{MN}}{2}$ & $\,M\!\cdot\!\min \left( {q,\frac{N}{2}} \right)$ & 5.3 dB & 10.6 dB & & $L+1$ & $ML$ & $L$ & -Sensetive to CEE. \\ 

\hline 
Successive CMF & Suc-CM & Sel-FW & & $\frac{{MN}}{2}$ & $\,M\! \cdot \! \min \left( {q,\frac{N}{2}} \right)$ & 4.6 dB & 4.8 dB & & $L+1$ & $ML$ & $L$ & -Robust against CEE,\newline -Added Complexity. \\ 

\hline 
\end{tabular} 
\label{tb:Compare}
\end{minipage}
\end{table*}


The first developed computing scheme is introduced in~\cite{BN1}, called standard computing (Std-CM) scheme, in which each single-antenna relay, independently of the other relays, decodes only one equation with the highest possible rate. This leads to an integer optimization problem in each relay to find the integer ECV of its equation.
In Std-CM scheme, the local channel state information (CSI)
should be known by the relay. 
Following~\cite{BN1}, different computing schemes have been suggested in the literature that can be employed as a part of the CMF methods over multi-user multi-relay networks, e.g.,~\cite{MIMO-CMF, myIET, Blind, IFLR, Lili, SIF-opt, SIF-arXiv, SumCap}.
In~\cite{P11, Azimi-IET, Nara5}, and~\cite{P01}, the Std-CM scheme is used  for different network structures.
In~\cite{MIMO-CMF}, the Std-CM scheme is generalized to the case of multiple-antenna users and relays.
In~\cite{myIET}, to reduce the optimization complexity, the authors propose the simplified version of Std-CM scheme that limits the integer ECVs to be selected from a predetermined set. 
In~\cite{Blind}, a blind computing scheme is introduced that, as opposed to other referred works, requires no CSI at the relays, at the cost of being sub-optimal.
In~\cite{Nara2}, a computing scheme is designed to address the timing asynchronism  in CMF methods.
The multi-input multi-output (MIMO) detection scheme suggested in~\cite{IFLR}, called integer-forcing linear receiver (IFLR), simultaneously recovers multiple ECVs and can be employed as a computing scheme by a multi-antenna relay. 
In the scheme utilized by~\cite{Lili} and~\cite{P01}, each relay finds a number of not-necessarily independent integer ECVs with the highest computation rates.

The performance of computing schemes can be improved by utilizing the previously decoded equations in decoding the subsequent equations~\cite{BN-SCMF}. The idea follows the same intuition as in successive interference cancellation (SIC) used in multi-user receivers~\cite[Ch. 7]{verdu} and is partially studied in several works~\cite{BN-SCMF, SIF-Zhan, SIF-opt, SIF-arXiv}. 
In~\cite{BN-SCMF}, SIC is implemented in a single-antenna relay for recovering two equations. 
In~\cite{SIF-Zhan}, a variant of IFLR scheme is proposed based on SIC. 
In~\cite{SIF-opt, SIF-arXiv}, the authors modify the IFLR scheme to take advantage of the remaining correlations among noises at the equalizer's output via noise-prediction, and improve the detection performance. In this approach, the receiver uses previous equations at each step, to reduce the effective noise in subsequent recovering steps.
This modified scheme, called successive integer-forcing, is a generalized form of the schemes in~\cite{BN-SCMF} and~\cite{SIF-Zhan}.

The first forwarding strategy, for CMF method, is  the one employed in original CMF method~\cite{BN1}, called standard forwarding (Std-FW) strategy. 
With the Std-FW, all $M$ relays send in turn their decoded equations to the destination.
The Std-FW strategy is also employed in~\cite{MIMO-CMF} and~\cite{myIET}.
In~\cite{Lili} and~\cite{P01}, a cooperative forwarding strategy is proposed to find linearly independent ECVs with highest possible rates. 
A centralized forwarding strategy is exploited in~\cite{P11}, where all relays send their ECVs along with their corresponding rates to the destination and the destination selects the relays with the highest computation rates that have linearly independent ECVs. 

One of the main challenges for CMF methods is the rank failure problem, in which the received equations by the destination may be linearly dependent, and hence the destination cannot recover its desired messages. This problem deteriorates the performance of CMF methods considerably and leads to a low order of diversity~\cite{myIET}. 
To decrease the probability of rank failure problem,~\cite{BN1} imposes a constraint on the selected integer ECVs in each relay, and~\cite{myIET} employs large number of relays. 
Also,~\cite{Lili,P11,P01}, suggest cooperation among relays or using a centralized coordinator. Although cooperative and centralized approaches decrease the probability of rank failure significantly, they require additional signaling overhead, feedback channels, or global CSI~\cite{Azh1,Azh2,Azh4}.
Assuming global CSI at the users,~\cite{P03},~\cite{Vite2}, and~\cite{P06} design transmit precoders for the users, to reduce the probability of selecting dependent equations by the relays.
However, none of the mentioned methods remove the rank failure problem completely within the practical constraints of the system.

In this paper, we propose a novel forwarding strategy, referred to as selection forwarding (Sel-FW) strategy, to combat the rank failure problem. This strategy can be used for arbitrary number of relays/users, and, in combination with a proper computing scheme, solves the rank failure problem. The Sel-FW strategy needs the minimum number of orthogonal relay-to-destination (R-D) channels (see Section~\ref{subsec:Sel-FW} for details). 
Also, the proposed Sel-FW strategy requires negligible signaling overhead or feedback channels.
 As a proper computing scheme for Sel-FW strategy, we extend the Std-CM scheme to decode multiple linearly independent equations in each relay. We refer to this technique as extended computing (Ext-CM). Moreover, to increase the computation rates of the equations at relays, we exploit the SIC idea in Ext-CM scheme, and develop the successive computing (Suc-CM) scheme. Employing the Suc-CM scheme leads to enhanced performance compared to the Ext-CM scheme. 

Considering the combinations of the Sel-FW strategy with the Ext-CM or Suc-CM schemes, we propose two novel CMF methods. First, we introduce the extended CMF method, which is the Sel-FW strategy in combination with the Ext-CM scheme. Second, we propose the successive CMF method that utilizes the Sel-FW strategy along with the Suc-CM scheme. Moreover, we consider the generalized version of the original CMF method~\cite{BN1} with multiple-antenna relays as the benchmark approach. The original CMF method consists of the Std-FW strategy and the Std-CM scheme (see Table~\ref{tb:Compare}).

In summary, the main contributions of our work, compared to the state-of-the-art schemes, can be outlined as:
\begin{enumerate}
\item We propose two novel CMF methods for the multi-user multi-relay networks, namely, extended CMF and successive CMF methods, with arbitrary number of users/relays and relay antennas. As opposed to aforementioned CMF methods, our proposed methods solve the rank failure problem, use the minimum required number of orthogonal R-D  channels,
 impose negligible signaling overhead or feedback channels to the network,
  and each relay requires only local CSI.
All these are gained at the cost of added complexity, compared to the original CMF method, due to finding larger number of ECVs at the relays. 
\item Our paper is different from~\cite{BN-SCMF, SIF-Zhan, SIF-opt, SIF-arXiv} because, first, we exploit the Suc-CM scheme in a multi-user multi-relay network. Second, we derive concise formulations (Equations ~(\ref{eq:bk_opt})-(\ref{eq:Q_k})) and explicit framework (in Section~\ref{subsec:Suc-CM}) for the Suc-CM scheme for general setups with different number of users/relay antennas. Furthermore, we introduce a novel approach to significantly facilitate the solution of the integer optimization problem in the Suc-CM scheme (see Theorem~\ref{th:Opt_sic}). 
\item We provide theoretical diversity analysis for our proposed methods in the cases with different R-D channels (Theorems~\ref{th:d-Ideal}, \ref{th:d-NF}, and~\ref{th:d-WF}). None of the derived analyses have been presented before.
\item As opposed to~\cite{BN1}, which is sensitive to channel estimation error (CEE)~\cite{CEE, myIET}, we show, through numerical simulations, that the Successive CMF is significantly robust against the CEE. This makes the successive CMF a proper method for practical applications. 
\end{enumerate}

Our diversity analysis and numerical simulations indicate that extended CMF and successive CMF methods achieve full diversity, i.e. the maximum possible diversity order, of the multi-user multi-relay network, provided that the R-D channels have a certain minimum quality. Furthermore, we show that the successive CMF method provides signal-to-noise ratio (SNR) gains and high robustness against CEE, compared to the extended CMF method, while has more complex structure. Both methods outperform the original CMF method considerably.

The rest of this paper is organized as follows. In Section~\ref{sec:model}, the system model is introduced. The forwarding strategies and computing schemes are presented in Section~\ref{sec:dest} and~\ref{sec:relay}, respectively. Closed-form equations and modified optimization problem for the successive case are given in Section~\ref{sec:relay}, as well. Section~\ref{sec:perf} includes the performance analysis of the proposed methods. Simulation and numerical results are presented in section~\ref{sec:simul}. Finally, section~\ref{sec:concl} concludes the paper.

\textbf{Notations:} 
Lower and upper boldfaced letters are used for column vectors and matrices, respectively. The symbol $\mathbf{I_n}$ stands for the $n \times n$ identity matrix. For a vector or matrix, $\left\| . \right\|$ and ${\left(  \cdot  \right)^T}$ indicate the Frobenius norm and  transpose operator, respectively. The operator $E \{ \cdot \}$ denotes the expectation operator. The notation ${\mathbf{x}} \perp\!\!\!\perp \left\{ {{\mathbf{y}},{\mathbf{z}}} \right\}$    indicates the linear independency of vector $\mathbf{x}$ and the set of vectors  $\left\{ {{\mathbf{y}},{\mathbf{z}}} \right\}$. The function $\log^+ {\left( x \right)}$ is equal to $\max \{ \log {\left(x\right)},0 \}$.

\section{System Model} \label{sec:model}
\begin{figure}[t]
\renewcommand{\figurename}{Fig.}
\centering
\includegraphics[trim = 5mm 2mm 0mm 0mm, clip, width =\Wf \columnwidth]{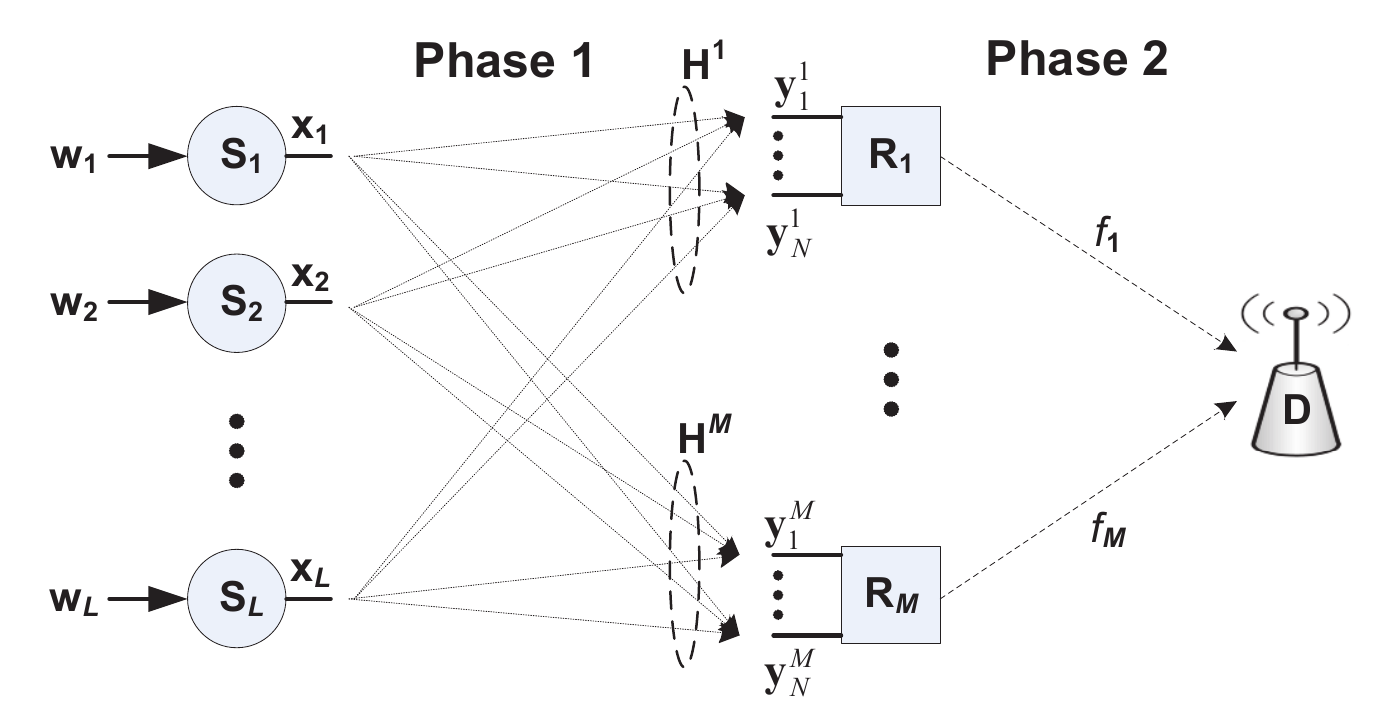}
\caption{\small System model.}
\label{fig:Model}
\end{figure}

We consider a network, shown in Fig.~\ref{fig:Model}, consisting of $L$ users, as the message sources, $M$ multi-antenna relays, and one common destination as the information sink. 
The users and the destination exploit a single antenna.
 Each relay is equipped with $N$ receive antennas. The network aims to reliably convey all messages from the sources to the destination with the highest possible rate. We assume that there is no direct link between the sources and the destination.

The real channel coefficient from source $l, l=1,\ldots,L$, to antenna $n, n=1,\ldots,N$, of the relay $m, m=1,\ldots,M$, is denoted by $h_{ln}^{m}$.
The $M$ channels from relays to the destination are orthogonal point-to-point channels. We consider two cases of ideal and non-ideal R-D channels. In the ideal case, the R-D channels are noiseless with sufficient capacity to transfer the required information without errors. For the non-ideal case, each R-D channel has the real coefficient $f_m, m=1,\ldots,M$, and independent zero-mean additive white Gaussian noise (AWGN) with variance ${\sigma'}_m^2, m=1,\ldots,M$. The relays are supposed to have the same power constraint $P_R$.

We define the channel coefficient matrix ${{\mathbf{H}}^{m}}$, corresponding to $L\times N$ MIMO channel from users to  relay $m$, as
\begin{equation} \label{eq:H_m}
{{\mathbf{H}}^{m}} = \left[ {{\mathbf{h}}_1^{m},{\mathbf{h}}_2^{m}, \ldots ,{\mathbf{h}}_N^{m}} \right], m=1,\ldots,M,
\end{equation}
where ${\mathbf{h}}_n^{m}$ is the channel coefficient vector corresponding to the links between different users and the $n$-th antenna of the relay $m$, as 
\begin{equation} \label{eq:hn_m}
{\mathbf{h}}_n^{m} = {\left[ {h_{1n}^{m},h_{2n}^{m}, \ldots ,h_{Ln}^{m}} \right]^{{T}}}, n=1,\ldots,N.
\end{equation}
 
Our system structure is based on the standard CMF, proposed in~\cite{BN1}. The user $l, l=1,\ldots,L$, exploits a lattice encoder  to map its corresponding message $\mathbf{w}_l$ to a real symbol $\mathbf{x}_l$ of length $t_0$, which is a lattice point with $\frac{1}{t_0} E{\left\| {{\mathbf{x}_l}} \right\|^2} = 1$~\cite{Nara1}. We assume that the power constraint of user $l$ is $P_l$. Thus, the user $l$ transmits the symbol ${\sqrt {{P_l}} {{\mathbf{x}}_l}}$ over the channel.

Each transmission frame consists of two phases. In the first phase, all users transmit their symbols simultaneously to the relays. Hence, the received signal at $n$-th antenna of relay $m$ can be expressed as
\begin{equation} \label{eq:yn_m}
{\mathbf{y}}_n^{m} = \sum_{l = 1}^L {h_{ln}^{m}\sqrt {{P_l}} {{\mathbf{x}}_l}}  + {\mathbf{n}}_n^{m}, m=1,\ldots,M, n=1,\ldots,N,
\end{equation}
where ${\mathbf{n}}_n^{m}$ is the zero-mean additive white Gaussian noise vector with variance $\sigma_{nm}^2$. 
In the second phase, the relays send their information through the orthogonal point-to-point channels (e.g. consecutive time slots) to the destination.

\subsection{Fading Channel} \label{subsec:channels}
As it is well-known, a channel coefficient in heavily built-up urban environments can be modeled as a zero-mean circularly-symmetric complex Gaussian random variable that results in Rayleigh and uniform distributions for gain and phase components of the channel, respectively~\cite[Ch. 4]{Rappaport}. Since in most scenarios (e.g. simple point-to-point channels), the phase can be compensated, the channel coefficients can be modeled as real Rayleigh variables. However, since in CMF, all users transmit simultaneously and each relay receives a linear combination of the transmitted signals, the channels phases cannot be compensated. Therefore, for the real case, i.e. when the transmitted signals are real, each relay can exploit the in-phase or quadrature component of its received signals. Hence, the effective channel is equal to the real or imaginary part of the complex channel, which can be modeled as a real zero-mean Gaussian random variable. The case in which each relay employs both the in-phase and quadrature components of its received signals, to achieve an improved performance, is equivalent to a relay that uses only one of the signals components and has twice the number of receive antennas.

For the complex case, each user sends two different messages over real and imaginary parts of the channels and each relay exploits both the in-phase and quadrature components of its received signals. For this case, in~\cite{BN1} it is shown that by real-valued decomposition of the complex channel matrix, the complex $L \times N$ MIMO channel between each user and each relay can equivalently be modeled as a real $2L \times 2N$ MIMO channel. Note that the SNR is the same for the real and complex models~\cite{BN1}.
Therefore, to cover all discussed cases, we assume that each relay considers only the in-phase components of its received signals. Hence, a real zero-mean Gaussian distribution with unit variance is considered for the channel $h_{ln}^{m}$  for all values of $l$, $m$, and $n$. Note that our assumptions are in harmony with~\cite{BN1} and~\cite{IFLR}, and the same procedure as in the paper is applicable in the cases with complex distributions.

For the case of non-ideal R-D channels, since we consider orthogonal point-to-point channels, the channel noises can be compensated, and hence the assumption of real channels is applicable. To cover different R-D channel conditions, we consider the Nakagami distribution with the parameter~$\mu$ for the channels coefficients. The fading severity decreases with $\mu$.
Rayleigh and AWGN channels are the special cases of the Nakagami channel corresponding to $\mu=1$ and $\mu = \infty$, respectively~\cite{Karbalayghareh}. The case of ideal R-D channels is equal to noiseless Nakagami channels with $\mu = \infty$.
From~\cite{Nakagami}, a $N_t \times N_r$ MIMO channel with Nakagami distributed channel coefficients and parameter $\mu$ can be modeled by a SISO channel with Nakagami distribution and parameter $q=\mu N_t N_r$. Hence, without loss of generality, we consider a single channel coefficient $f_m ,m=1,\ldots,M$ for each R-D channel that follows the independent real Nakagami distribution with parameter $q$ and unit variance.
Furthermore, we assume block-fading conditions where the channel coefficients are constant during each transmission frame and independent of the ones in the other transmission frames.

\section{Forwarding Strategies} \label{sec:dest}
Relay $m, m=1,\ldots,M$, receives $N$ signals ${\mathbf{y}}_n^{m}, n=1,\ldots,N$, as expressed in~(\ref{eq:yn_m}), in the first transmission phase. By exploiting its $N$ received signals, the relay aims to compute an equation. An equation $\mathbf{u}$ is a linear combination of the users' symbols $\mathbf{x}_l, l=1,\ldots,L$, with integer coefficients, as:
\begin{equation} \label{eq:u}
{\mathbf{u}} = \sum_{l = 1}^L {{a_l}{{\mathbf{x}}_l}},
\end{equation}
where ${\mathbf{a}} = {\left[ {{a_1},{a_2}, \ldots ,{a_L}} \right]^{{T}}} \in {\mathbb{Z}^L}$ is referred to as the equation coefficient vector (ECV) corresponding to  equation $\mathbf{u}$. 
The rate of decoding an equation is called computation rate. Note that a set of equations are called linearly independent if and only if their corresponding ECVs (equation coefficients vectors) are linearly independent. Moreover, the rate of recovering a message (or an equation) from a set of equations is equal to the minimum computation rate of the equations that are used in its recovery.

The forwarding strategy is an important part of the CMF methods that determines how the information, including selected ECVs, decoded equations, and cooperation and feedback signals, flows over the network. In the following, we discuss the forwarding strategies.

\subsection{Standard Forwarding (Std-FW) Strategy}
Std-FW is the forwarding strategy of the original CMF method~\cite{BN1}. 
However, as original CMF is used as a benchmark for comparisons of our proposed methods, it is described here.  
In the original CMF method, each relay decodes the best equation, i.e. the equation with the highest computation rate, based on its received signals (see Section~\ref{subsec:relay:stdcmf}). In the second transmission phase, each relay sends its decoded equation to the destination in its dedicated channel. Hence, the destination receives $M$ equations from the relays. The destination, selects  $L$ linearly independent equations, with the highest computation rates, out of the $M$ received equations, and then can  recover all  messages. 

In the original CMF method~\cite{BN1}, since each relay finds its best equation statistically independently of the other relays, the equations received by the destination may be linearly dependent. If the received equations are linearly dependent, the coefficient matrix of the equations is singular and the rank failure  occurs. Thus, the destination cannot recover all  messages. Rank failure problem results in  significant performance degradation in original CMF method and decreases the diversity order drastically~\cite{myIET}.

In the Std-FW strategy, all $M$ relays send in turn their decoded equations to the destination. Hence, it requires $M$ time-slots or orthogonal channels. In the Std-FW strategy, the number of relays should be equal or greater than the number of users.

\subsection{Selection Forwarding (Sel-FW) Strategy} \label{subsec:Sel-FW}
Our proposed forwarding strategy, i.e. Sel-FW, is used in both of our extended CMF and successive CMF methods. In these methods, each relay finds the $L$ best ECVs, i.e. the $L$ linearly independent ECVs with the highest computation rates, based on its received signals (see Section~\ref{sec:relay}). Let $\rho_m$ denote the minimum of the computation rates for the best ECVs of relay $m$. Note that from the best equations of relay $m$ all  $L$ messages can be recovered with rate $\rho_m$.
In Sel-FW strategy, the relay with highest $\rho_m$ is selected. 
The relay selection can be performed in either of the following ways:

\begin{itemize}
\item Each relay sends its rate $\rho_m$ to the destination. Destination selects the highest rate and informs the selected relay through a low-rate feedback channel. The feedback rate is $\lceil \log_2{M} \rceil$ bits per relay selection interval.
\item Similar to~\cite{R22}, each relay sets a timer with the value $T_m$ proportional to the inverse of its rate $\rho_m$. The timers start to count down at the beginning of the second transmission phase. The relay whose timer reaches zero first (which has the highest rate) broadcasts a flag to inform other relays and is selected as the best relay. We assume that the flag is a short-time high-energy signal that can be sensed by the other relays with a probability close to one. This approach needs no feedback channel.
\end{itemize}

Note that the relay selection is necessary once the channels coefficients have changed, i.e. the best relay is fixed for the coherence interval of the channels. Thus, for slow fading channels, relay selection imposes negligible additional complexity and overhead to the network.

The selected relay decodes and forwards its $L$ best equations, corresponding to its best ECVs, to the destination through $L$ orthogonal channels. Utilizing the $L$ received independent equations, the destination can solve and recover all messages, without encountering the rank failure problem. Note that, since $L$ messages are transmitted over the network, the destinations needs at least $L$ equations from the relays to recover all  messages. Hence, the minimum number of required R-D orthogonal channels is equal to $L$, that is achieved by Sel-FW strategy.

Note that, although selecting one relay is not globally optimal for recovering $L$ independent equations with highest rates, it eliminates the need for information exchange among the relays, as it is required in cooperative strategies.

\begin{remark} \label{Re:Complexity}
Based on the Sel-FW strategy, each of the $M$ relays finds $L$ ECVs, but only the selected relay (with highest $\rho_m$) decodes the $L$ equations corresponding to its ECVs. On the other hand, in the original CMF method, each of the $M$ relays selects an ECV and decodes the corresponding equation. Hence, to compare the overall computational complexity, in the former, $ML$ ECVs are selected and $L$ equations are decoded, while in the latter $M$ ECVs are selected and $M$ equations are decoded.
\end{remark}

\section{Computing Schemes} \label{sec:relay}
The Std-FW and Sel-FW strategies require computing schemes that can be employed in a relay to find the best ECV or the best $L$ linearly independent ECVs, respectively. Moreover, the technique of computing equations, corresponding to the selected ECVs, should be specified by the computing schemes. 

We assume that each relay knows only local CSI, as in, e.g.,~\cite{BN1}. All equations of the current section are written for each relay $m, m=1,\ldots,M$. Hence, for simplicity of presentation, we remove the index $m$ from the equations.

\subsection{Standard Computing (Std-CM) Scheme} \label{subsec:relay:stdcmf}
A relay receives $N$ signals, $\mathbf{y}_n, n=1,\ldots,N$, at its antennas in the first transmission phase, as expressed in~(\ref{eq:yn_m}). The normalized received signal at the $n$-th antenna can be written as
\begin{equation} \label{eq:ytilde}
{{\mathbf{\tilde y}}_n} \triangleq \frac{{{{\mathbf{y}}_n}}}{{{\sigma _n}}} = \sum\limits_{l = 1}^L {{g_{ln}}{{\mathbf{x}}_l}}  + {{\mathbf{z}}_n},
\end{equation}
where ${g_{ln}} = {h_{ln}}\frac{{\sqrt {{P_l}} }}{{{\sigma _n}}}$ is the instantaneous received SNR at $n$-th antenna of the relay from user $l$, and ${{\mathbf{z}}_n} = \frac{{{{\mathbf{n}}_n}}}{{{\sigma _n}}}$ is the received noise with unit variance.
Let us define the vector  ${{\mathbf{g}}_n}$ and matrix $\mathbf{G}$ as
\begin{equation} \label{eq:g_n}
{{\mathbf{g}}_n} \triangleq {\left[ {{g_{n1}}, \ldots ,{g_{nL}}} \right]^{{T}}},
\end{equation}
and
\begin{equation} \label{eq:G}
     {\mathbf{G}} = {\left[ {{{\mathbf{g}}_1}, \ldots ,{{\mathbf{g}}_N}} \right]_{L \times N}}.
\end{equation}
To recover an equation $\mathbf{u}$ with ECV ${\mathbf{a}} = {\left[ {{a_1},{a_2}, \ldots ,{a_L}} \right]^{{T}}} \in {\mathbb{Z}^L}$, as expressed in~(\ref{eq:u}), the relay combines the normalized received signals with coefficient vector ${{\mathbf{b}}} = {\left[ {{b_{1}}, \ldots ,{b_{N}}} \right]^{{T}}\in {\mathbb{R}^N}}$, as
\begin{equation} \label{eq:ybar}
{{\mathbf{\bar y}}} = \sum_{n = 1}^N {{b_{n}}{{{\mathbf{\tilde y}}}_n}}.
\end{equation}
Thus, the equivalent channel from the users to the relay is modeled as
\begin{equation} \label{eq:ybar2}
{{\mathbf{\bar y}}} = \sum_{l = 1}^L {{{\bar g}_{l}}} {{\mathbf{x}}_l} + {{\mathbf{\bar z}}},
\end{equation}
where the equivalent noise ${{\mathbf{\bar z}}}$ is a zero-mean AWGN with variance ${\left\| {{{\mathbf{b}}}} \right\|^2}$, and the equivalent channel vector ${{\mathbf{\bar g}}} = {\left[ {{{\bar g}_{1}}, \ldots ,{{\bar g}_{L}}} \right]^{{T}}}$ is calculated as
\begin{equation} \label{eq:g_bar}
{{\mathbf{\bar g}}} = {\mathbf{G}}{{\mathbf{b}}}.
\end{equation}
From~\cite{MIMO-CMF}, the computation rate of the equation with ECV ${\mathbf{a}} = {\left[ {{a_1},{a_2}, \ldots ,{a_L}} \right]^{{T}}} \in \mathbb{Z}^L$ is equal to
\begin{equation} \label{eq:R_G_b_a}
R({\mathbf{G}},{{\mathbf{b}}},{\mathbf{a}}) = \frac{1}{2}{\log ^ + }\left( {\frac{1}{{{{\left\| {{{\mathbf{b}}}} \right\|}^2} + {{\left\| {{\mathbf{G}}{{\mathbf{b}}} - {\mathbf{a}}} \right\|}^2}}}} \right),
\end{equation}
and the optimum value of $\mathbf{b}$ maximizing~(\ref{eq:R_G_b_a}) is given by
\begin{equation} \label{eq:b}
{{\mathbf{b}}} = {\left( {{\mathbf{I}_N} + {{\mathbf{G}}^T}{\mathbf{G}}} \right)^{ - 1}}{{\mathbf{G}}^{T}}{\mathbf{a}}.
\end{equation}
Substituting~(\ref{eq:b}) into~(\ref{eq:R_G_b_a}) yields the maximum achievable computation rate of the relay for decoding an equation with ECV $\mathbf{a}$, as
 \begin{equation}\label{eq:R_G_a}
 R({\mathbf{G}},{\mathbf{a}}) = \frac{1}{2}{\log ^ + }\left( {\frac{1}{{{{\mathbf{a}}^{\mathrm{T}}}{\mathbf{Qa}}}}} \right),
 \end{equation}
where
\begin{equation} \label{eq:Q}
{\mathbf{Q}} = {\left( {{\mathbf{I}}_L + {\mathbf{G}}{{\mathbf{G}}^{T}}} \right)^{ - 1}},
\end{equation}
is a positive definite matrix.
Therefore, the relay can find the best ECV from the following optimization problem
\begin{equation} \label{eq:a1}
\begin{aligned}
 {{\mathbf{a}}_1} &= \arg \max_{{\mathbf{a}} \in {\mathbb{Z}^L},{\mathbf{a}} \ne {\mathbf{0}}} {R\left({\mathbf{G}},{\mathbf{a}}\right)}  \\
   &= \arg \max_{{\mathbf{a}} \in {\mathbb{Z}^L},{\mathbf{a}} \ne {\mathbf{0}}} {\frac{1}{2}{\log ^ + }\left( {\frac{1}{{{{\mathbf{a}}^{{T}}}{\mathbf{Qa}}}}} \right)}  \\
   &= \arg \min_{{\mathbf{a}} \in {\mathbb{Z}^L},{\mathbf{a}} \ne {\mathbf{0}}} {{{\mathbf{a}}^{{T}}}{\mathbf{Qa}}},
\end{aligned}
\end{equation}
where the last equality holds since $\log^+(\cdot)$ is a monotonically increasing function.

The above integer optimization problem is equivalent to the shortest vector problem (SVP), and has no closed-form solution~\cite{P05,myIET}. Different approaches can be applied to calculate the optimum ECV numerically~\cite{Fin, Lili, P09 , P10, Algebraic,Soussi,P07}. Since $\mathbf{Q}$ is a positive definite matrix, we can follow the same method as in~\cite{Fin} to find the optimal vectors.

To decode (compute) the equation corresponding to optimum ECV $\mathbf{a}_1$, the relay calculates the combining coefficient vector $\mathbf{b}_1$ from~(\ref{eq:b}) and generates $\mathbf{\bar y}_1$ from~(\ref{eq:ybar}). Then, the signal $\mathbf{\bar y}_1$ is used as the input of the relay lattice decoder to recover the desired equation ${{\mathbf{\hat u}}_1}$.

The Std-CM scheme presented above, is the generalized (with multiple antenna) version of  the computing scheme of the original CMF method~\cite{BN1} and is the basis of the other  presented computing schemes in this paper.

\subsection{Extended Computing (Ext-CM) Scheme} \label{subsec:Ext-CM}
Ext-CM is used as the computing scheme of our extended CMF method. In our proposed forwarding strategy, i.e. Sel-FW, the relay needs to find $L$ independent ECVs with the highest computation rates. We extend the Std-CM scheme to find the desired ECVs as follows. 

The relay finds the first ECV from~(\ref{eq:a1}). The other ECVs are calculated sequentially through
\begin{equation} \label{eq:ak}
{{\mathbf{a}}_k}= \arg \min_{\substack{{\mathbf{a}} \in {\mathbb{Z}^L},{\mathbf{a}} \ne {\mathbf{0}}\\
{\mathbf{a}} \perp\!\!\!\perp \left\{ {{{\mathbf{a}}_1}, \ldots ,{{\mathbf{a}}_{k - 1}}} \right\}}}
 {{{\mathbf{a}}^{{T}}}{\mathbf{Q} \mathbf{a}}}, \quad k=2,\ldots,L,
\end{equation}
where the constraint ${\mathbf{a}} \perp\!\!\!\perp \left\{ {{{\mathbf{a}}_1}, \ldots ,{{\mathbf{a}}_{k - 1}}} \right\}$ guarantees that the ECV $\mathbf{a}_k$ is linearly independent of the previous ECVs $\mathbf{a}_1,\ldots,\mathbf{a}_{k-1}$.

The above optimization can be solved by employing the same method as in~\cite{Fin}, except that the search space is over vectors independent of the  previous ECVs. Moreover, the approaches suggested in~\cite{P05, P07,Lili,Algebraic,P08} can be exploited to find the solution of~(\ref{eq:ak}) with reduced complexity.

By finding the ECVs, the relay can decode the corresponding equations. Specifically, for decoding the $k$-th equation, it calculates the combining coefficient vector $\mathbf{b}_k$ from~(\ref{eq:b}) and generates the combined signal $\mathbf{\bar y}_k$ from~(\ref{eq:ybar}). Then, the signal $\mathbf{\bar y}_k$ is used as the input of the relay lattice decoder to recover the $k$-th equation ${{\mathbf{\hat u}}_k}$. From~(\ref{eq:R_G_a}) and using the Ext-CM scheme, the computation rate of  equation ${{\mathbf{\hat u}}_k}$ with ECV $\mathbf{a}_k$ is given by
\begin{equation} \label{eq:Rstd_k}
R^{\mathrm{(Ext)}}_k \left( \mathbf{G} \right)= \frac{1}{2}{\log ^ + }\left( {\frac{1}{{{{\mathbf{a}_k}^{\mathrm{T}}}{\mathbf{Q}\mathbf{a}_k}}}} \right).
\end{equation}
Note that $R^{\mathrm{(Ext)}}_1 \left( \mathbf{G} \right) \geq \cdots  \geq R^{\mathrm{(Ext)}}_L \left( \mathbf{G} \right) $. Hence, the sum rate of recovering all  messages from these $L$ equations is equal to 
\begin{equation} \label{eq:Rsum-Ext}
R^{\mathrm{(Ext)}}_{\text{sum}}=L \times R^{\mathrm{(Ext)}}_L \left( \mathbf{G} \right).
\end{equation}

In Ext-CM scheme, $L$ linearly independent equations are selected sequentially, as expressed in~(\ref{eq:ak}). A different  approach, named IFLR, is presented in~\cite{IFLR} to recover multiple equations at the receiver of a point-to-point MIMO channel. However, since in IFLR the data streams of transmit antennas are independently coded, the scheme can be exploited in a multi-user scenario as well. The IFLR scheme results in the following optimization 
\begin{equation} \label{eq:Rsum-IFLR}
R^{\text{(IFLR)}}_{\text{sum}}=L \times \max_{\substack{ \Omega \subseteq {\mathbb{Z}^L}\setminus\mathbf{0},\,\, \left| \Omega \right|=L \\ \mathrm{rank}(\Omega)=L  }}  { \min_{{\mathbf{a}} \in \Omega }\,\, {\frac{1}{2}{\log ^ + }\left( {\frac{1}{{{{\mathbf{a}}^{{T}}}{\mathbf{Qa}}}}} \right)} },
\end{equation}
where ${\mathbb{Z}^L}\setminus\mathbf{0}$ denotes ${\mathbb{Z}^L}$ excluding the zero vector, $\left| \Omega \right|$ shows the cardinality of the set $\Omega$, and $\mathrm{rank}(\Omega)$ equals the rank of the matrix whose columns are the elements of $\Omega$. 
Comparing~(\ref{eq:Rsum-IFLR}) with~(\ref{eq:ak}) reveals that in IFLR scheme, all equations are selected simultaneously (jointly) to maximize the sum  rate, as opposed to our Ext-CM approach with sequential ECVs selection.
The IFLR scheme is shown to be optimal in terms of rates of the recovered equations~\cite{IFLR}. In the following theorem, we prove that our sequential selection of ECVs, achieves the same rate as the IFLR scheme, and hence, is optimal.

\begin{theorem} \label{th:IFLR}
For any given channel matrix $\mathbf{G}$, defined in~(\ref{eq:G}), the Ext-CM scheme achieves the same sum rate as the optimal IFLR scheme. 
\end{theorem}
\begin{IEEEproof}
See Appendix~\ref{ap:IFLR}.
\end{IEEEproof}

It is clear that Ext-CM has a lower complexity than the IFLR due to sequential, rather than joint, selection of ECVs, while providing the same performance as IFLR. As apposed to~\cite{IFLR} that uses the IFLR scheme in a point-to-point MIMO channel, we exploit the Ext-CM scheme in combination with Sel-FW strategy to form extended CMF method for multi-user multi-relay networks.

\subsection{Successive Computing (Suc-CM) Scheme} \label{subsec:Suc-CM}
Suc-CM is employed as the computing scheme of our successive CMF method. To find the best $L$ linearly independent ECVS and decoding  their corresponding equations, as required by Sel-FW strategy, we propose the Suc-CM scheme. In Ext-CM scheme, since all  $L$ equations are recovered from a single relay, the minimum rate of them, i.e. $R'_L$, tends to zero as $L$ increases. To solve this issue, we can use the previous decoded equations to improve the computation rates of the subsequent equations. 
  
In the first step of Suc-CM scheme, the first ECV $\mathbf{a}_1$ and its corresponding equation $\mathbf{\hat u}_1$ are determined the same as in the Std-CM and Ext-CM schemes. In the $k$-th step, $k=2,\ldots,L$, the relay desires to find the $k$-th ECV ${{\mathbf{a}}_k}$ and decode its corresponding equation ${{\mathbf{\hat u}}_k}$. From the previous steps, the linearly independent ECVs ${{\mathbf{a}}_1}, \ldots ,{{\mathbf{a}}_{k - 1}}$ and the equations ${{\mathbf{\hat u}}_1}, \ldots ,{{\mathbf{\hat u}}_{k - 1}}$ are known. Thus, by combining its antenna signals as well as the previously decoded equations, the relay generates the following signal
\begin{equation} \label{eq:ybar_sic}
{{\mathbf{\bar y }}_k} = \sum\limits_{n = 1}^N {{b_{kn}}{{{\mathbf{\tilde y}}}_n}}  + \sum\limits_{j = 1}^{k - 1} {{\beta _{kj}}{{{\mathbf{\hat u}}}_j}}.
\end{equation}
Here, ${{{\mathbf{\tilde y}}}_n}$ is the normalized signal defined in~(\ref{eq:ytilde}) and  ${{\mathbf{\hat u}}}_j$ is the decoded equation with ECV ${{\mathbf{a}}}_j$ in the $j$-th step that is expressed as
\begin{equation} \label{eq:uhat_j}
{\mathbf{\hat u}_j} = \sum_{l = 1}^L {{a_{jl}}{{\mathbf{x}}_l}}.
\end{equation}
The vectors ${{\mathbf{b}}_k} = {\left[ {{b_{k1}}, \ldots ,{b_{kN}}} \right]^{{T}}\in {\mathbb{R}^N}}$ and ${{\pmb{\beta }}_k} = {\left[ {\beta _1^{(k)}, \ldots ,\beta _{k - 1}^{(k)}} \right]^{{T}}}\in {\mathbb{R}^{k-1}}$ include the combination coefficients for the normalized antenna signals and the previous equations, respectively. Thus, the equivalent channel from the users to the relay is modeled as
\begin{equation} \label{eq:ybar2_sic}
{{\mathbf{\bar y}}_k} = \sum_{l = 1}^L {{{{\bar g}_{kl}}} {{\mathbf{x}}_l}} + {{\mathbf{\bar z}}_k},
\end{equation}
where the equivalent noise ${{\mathbf{\bar z}_k}}$ is a zero-mean AWGN with variance ${\left\| {{{\mathbf{b}}_k}} \right\|^2}$, and the equivalent channel vector ${{\mathbf{\bar g}}_k} = {\left[ {{{\bar g}_{k1}}, \ldots ,{{\bar g}_{kL}}} \right]^{{T}}}$ is calculated as
\begin{equation} \label{eq:g_bar_sic}
{{\mathbf{\bar g}}_k} = {\mathbf{G}}{{\mathbf{b}}_k} + {{\mathbf{A}}_{k - 1}}{{\pmb{\beta }}_k},
\end{equation}
where $\mathbf{G}$ is defined in~(\ref{eq:G}) and ${\mathbf{A}}_{k - 1}$ is a ${L \times ( {k - 1} )}$ matrix defined as
\begin{equation} \label{eq:A_k-1}
{{\mathbf{A}}_{k - 1}} = {\left[ {{{\mathbf{a}}_1}, \ldots ,{{\mathbf{a}}_{k - 1}}} \right]}.
\end{equation}

Similar to~(\ref{eq:R_G_b_a}), the computation rate of an equation with ECV ${\mathbf{a}} = {\left[ {{a_1},{a_2}, \ldots ,{a_L}} \right]^{{T}}}$ is
\begin{equation} \label{eq:R_G_b_a_sic}
R({\mathbf{G}},{{\mathbf{b}}_k},{{\pmb{\beta }}_k},{\mathbf{a}}) = \frac{1}{2}{\log ^ + }\!\! \left( \!{\frac{1}{{{{\left\| {{{\mathbf{b}}_k}} \right\|}^2}\! + {{\left\| {{\mathbf{G}}{{\mathbf{b}}_k}\! +\! {{\mathbf{A}}_{k - 1}}{{\pmb{\beta }}_k}\! -\! {\mathbf{a}}} \right\|}^2}}}}\! \right).
\end{equation}

It is worth noting that there are two terms in the denominator of the above equation reducing the rate. The first, i.e. ${{\left\| {{{\mathbf{b}}_k}} \right\|}^2}$, is related to the noise, and the second is due to mismatch between the equivalent channel coefficients ${{\mathbf{\bar g}}_k}$ in~(\ref{eq:g_bar_sic}) and the desired ECV $\mathbf{a} \in \mathbb{Z}^L$. The coefficient vectors $\mathbf{b}_k$ and $\pmb{\beta}_k$ can be adjusted to decrease the mismatch term. Comparing the mismatch terms in~(\ref{eq:R_G_b_a}) and~(\ref{eq:R_G_b_a_sic}) indicates that exploiting the previously decoded equations in successive CMF provides more degrees of freedom to reduce the mismatch, which  leads to  rate enhancement. Furthermore, utilizing larger number of antennas, i.e. $N$, increases the dimension of $\mathbf{b}_k$, resulting in similar effect, i.e. the received signals from  antennas can be combined to reduce the mismatch term.

To maximize the computation rate in~(\ref{eq:R_G_b_a_sic}), optimum values of $\mathbf{b}_k$ and $\pmb{\beta}_k$ as well as the corresponding maximum computation rate can be determined from the following theorem and corollary. These results are of interest because they are expressed in concise and closed forms.

\begin{theorem} \label{th:b_opt}
For a given ECV $\mathbf{a}$, the optimum ${{\mathbf{b}}_k}$ and ${{\pmb{\beta }}_k}$ that maximize the computation rate in~(\ref{eq:R_G_b_a_sic}) are 
\begin{align} 
{{\mathbf{b}}_k} &= {\left( {{\mathbf{I}_N} + {\mathbf{\tilde G}}_k^T{{{\mathbf{\tilde G}}}_k}} \right)^{ - 1}}{\mathbf{\tilde G}}_k^T{\mathbf{a}}, \label{eq:bk_opt}\\
 {{\pmb{\beta }}_k} &= {\left( {{\mathbf{A}}_{k - 1}^T{{\mathbf{A}}_{k - 1}}} \right)^{ - 1}}\!{\mathbf{A}}_{k - 1}^T\! \left(\! {{\mathbf{I}_L} \! - \! {\mathbf{G}}{{\left( {{\mathbf{I}_N}\! +\! {\mathbf{\tilde G}}_k^T{{{\mathbf{\tilde G}}}_k}} \right)}^{\! - 1}}\!{\mathbf{\tilde G}}_k^T} \! \right)\!{\mathbf{a}}, \label{eq:betak_opt}
\end{align}
where
\begin{align} 
&{{{\mathbf{\tilde G}}}_k} \triangleq {{\mathbf{F}}_{k - 1}}{\mathbf{G}},  \label{eq:Gtilde_k}\\
&{{\mathbf{F}}_{k - 1}} \triangleq \left( {{\mathbf{I}_L} - {{\mathbf{A}}_{k - 1}}{{\left( {{\mathbf{A}}_{k - 1}^T{{\mathbf{A}}_{k - 1}}} \right)}^{ - 1}}{\mathbf{A}}_{k - 1}^T} \right), \label{eq:F_k-1}
\end{align}
and $\mathbf{F}_0\triangleq \mathbf{I}_L,\,\,\,{{\pmb{\beta }}_1}\triangleq \left[ \, \right]$ .

\end{theorem}
\begin{IEEEproof}
See Appendix~\ref{ap:bk_opt_proof}.
\end{IEEEproof}

\begin{remark} \label{Re:F_k-1}
The ${{\mathbf{F}}_{k - 1}}$ is the matrix of projection onto the orthogonal complement of the subspace spanned by vectors  $\left\{ {{{\mathbf{a}}_1}, \ldots ,{{\mathbf{a}}_{k - 1}}} \right\}$~\cite[Ch. 5]{Meyer}. Hence,  it is an idempotent matrix, i.e.  ${\mathbf{F}}_{k - 1}^2 = {{\mathbf{F}}_{k - 1}}$. Moreover, we have ${\mathbf{F}}_{k - 1}^T = {{\mathbf{F}}_{k - 1}}$. 
\end{remark}

\begin{corollary} \label{cor:Ropt}
For a given ECV $\mathbf{a}$, the optimum computation rate of~(\ref{eq:R_G_b_a_sic}) is equal to
\begin{equation} \label{eq:R_G_a_sic}
R_k \left({\mathbf{G}},{\mathbf{a}}\right) = \frac{1}{2}{\log ^ + }\left( {\frac{1}{{{{\mathbf{a}}^T}{{{\mathbf{Q}}}_k}{\mathbf{a}}}}} \right),
\end{equation}
where
\begin{equation}\label{eq:Q_k}
{{\mathbf{Q}}_k} = {\mathbf{F}}_{k - 1}^T{\left( {{\mathbf{I}_L} + {{{\mathbf{\tilde G}}}_k}{\mathbf{\tilde G}}_k^T} \right)^{ - 1}}{{\mathbf{F}}_{k - 1}},
\end{equation}
and matrices ${{{\mathbf{\tilde G}}}_k}$ and ${\mathbf{F}}_{k - 1}$ are defined in~(\ref{eq:Gtilde_k}) and~(\ref{eq:F_k-1}), respectively.
\end{corollary}
\begin{IEEEproof}
By substitution of~(\ref{eq:bk_opt}) and~(\ref{eq:betak_opt}) in~(\ref{eq:R_G_b_a_sic}), and using the properties of ${\mathbf{F}}_{k - 1}$ in Remark~\ref{Re:F_k-1}, we get the desired results~(\ref{eq:R_G_a_sic}) and~(\ref{eq:Q_k}).
\end{IEEEproof}
\begin{lemma} \label{lem:Q_k}
The matrix $\mathbf{Q}_k$ in~(\ref{eq:Q_k}) is a positive semi-definite (and not a positive definite) matrix.
\end{lemma}
\begin{IEEEproof}
For any ${\mathbf{x}} \ne {\mathbf{0}}$ in $\mathbb{R}^L$, we have
\begin{equation}
{{\mathbf{x}}^T}\left( {{\mathbf{I}_L} + {{{\mathbf{\tilde G}}}_k}{\mathbf{\tilde G}}_k^T} \right){\mathbf{x}} = {\left\| {\mathbf{x}} \right\|^2} + {\left\| {{\mathbf{\tilde G}}_k^T{\mathbf{x}}} \right\|^2} > 0.
\end{equation}
Thus, the matrix $\left( {{\mathbf{I}_L} + {{{\mathbf{\tilde G}}}_k}{\mathbf{\tilde G}}_k^T} \right)$ is a positive definite matrix. As a result, its inverse is also a positive definite matrix and has a Cholesky decomposition of the form $\mathbf{L}\mathbf{L}^T$, where $\mathbf{L}$ is a lower triangular matrix with positive diagonal entries~\cite[Ch. 7]{Meyer}. For any ${\mathbf{x}} \ne {\mathbf{0}}$ in $\mathbb{R}^L$, we can write
\begin{equation}
\begin{aligned}
{{\mathbf{x}}^T}{{{\mathbf{Q}}}_k}{\mathbf{x}} &= {{\mathbf{x}}^T}{\mathbf{F}}_{k - 1}^T{\left( {{\mathbf{I}_L} + {{{\mathbf{\tilde G}}}_k}{\mathbf{\tilde G}}_k^T} \right)^{ - 1}}{{\mathbf{F}}_{k - 1}}{\mathbf{x}} \\
 &= {{\mathbf{x}}^T}{\mathbf{F}}_{k - 1}^T{\mathbf{L}}{{\mathbf{L}}^T}{{\mathbf{F}}_{k - 1}}{\mathbf{x}}\\
  &= {\left\| {{{\mathbf{L}}^T}{{\mathbf{F}}_{k - 1}}{\mathbf{x}}} \right\|^2} \geqslant 0.
\end{aligned}
\end{equation}
Hence, $\mathbf{Q}_k$ is a positive semi-definite matrix. Now, consider the case ${\mathbf{x}} = {{\mathbf{a}}_1} \ne {\mathbf{0}}$. Since ${{\mathbf{a}}_1}$ is in the span of columns of $\mathbf{A}_{k-1}$, based on Definition~\ref{def:complement} in Appendix~\ref{ap:definitions}, we have that the vector ${{\mathbf{a}}_1}$ has no components in orthogonal complement of the span of columns of $\mathbf{A}_{k-1}$. Therefore, from definition of ${{\mathbf{F}}_{k - 1}}$ in Remark~\ref{Re:F_k-1},  we get ${{\mathbf{F}}_{k - 1}}{\mathbf{a}_1} = \mathbf{0}$ and ${{\mathbf{a}_1}^T}{{{\mathbf{Q}}}_k}{\mathbf{a}_1}=0$. Thus, ${{{\mathbf{Q}}}_k}$ is not a positive definite matrix.
\end{IEEEproof}

From~(\ref{eq:R_G_a_sic}) in Corollary~\ref{cor:Ropt}, the relay, in step $k$, can find the best ECV ${{\mathbf{a}}_k}$ from the following optimization
\begin{equation} \label{eq:ak_sic}
\begin{aligned}
 {{\mathbf{a}}_k} &= \arg \max_{\substack{{\mathbf{a}} \in {\mathbb{Z}^L},{\mathbf{a}} \ne {\mathbf{0}}\\
{\mathbf{a}} \perp\!\!\!\perp \left\{ {{{\mathbf{a}}_1}, \ldots ,{{\mathbf{a}}_{k - 1}}} \right\}}} {R_k \left({\mathbf{G}},{\mathbf{a}}\right)}  \\
   &= \arg \min_{\substack{{\mathbf{a}} \in {\mathbb{Z}^L},{\mathbf{a}} \ne {\mathbf{0}}\\
{\mathbf{a}} \perp\!\!\!\perp \left\{ {{{\mathbf{a}}_1}, \ldots ,{{\mathbf{a}}_{k - 1}}} \right\}}} {{{\mathbf{a}}^{{T}}}{\mathbf{Q}_k \mathbf{a}}},
\end{aligned}
\end{equation}
where the matrix $\mathbf{Q}_k$ is defined in~(\ref{eq:Q_k}). The constraint ${\mathbf{a}} \perp\!\!\!\perp \left\{ {{{\mathbf{a}}_1}, \ldots ,{{\mathbf{a}}_{k - 1}}} \right\}$ guarantees that the ECV $\mathbf{a}_k$ is linearly independent of the previous ECVs $\mathbf{a}_1,\ldots,\mathbf{a}_{k-1}$.

In Lemma~\ref{lem:Q_k}, it is proved that the matrix ${{\mathbf{Q}}_k}$ is not a positive definite matrix. In this case, efficient methods such as the one in~\cite{Fin}  cannot be employed, and finding the solution of the optimization problem~(\ref{eq:ak_sic}) will be time-consuming. In Section~\ref{subsec:Suc-opt}, we propose an approach to overcome this issue.

By finding ECV  $\mathbf{a}_k$ from~(\ref{eq:ak_sic}), the relay decodes the corresponding equations ${{\mathbf{\hat u}}_k}$ in step $k$ as follows.
The relay calculates the combining coefficient vectors $\mathbf{b}_k$ and $\pmb{\beta}_k$ from~(\ref{eq:bk_opt}) and~(\ref{eq:betak_opt}), respectively, to generate the combined signal $\mathbf{\bar y}_k$ as in~(\ref{eq:ybar_sic}). Then, the signal $\mathbf{\bar y}_k$ is used by the  lattice decoder to recover the $k$-th equation ${{\mathbf{\hat u}}_k}$.

Since in the Suc-CM scheme, the previous equations are used for decoding the current equation ${{\mathbf{\hat u}}_k}$, the computation rate of the ${{\mathbf{\hat u}}_k}$ with ECV $\mathbf{a}_k$ is obtained as
\begin{equation} \label{eq:Rsec_k}
\begin{aligned}
R_k^{{\mathrm{(Suc)}}}({\mathbf{G}}) &= \,\mathop {\min }\limits_{1 \leqslant j \leqslant k} \, \, \,{R_j}({\mathbf{G}},{{\mathbf{a}}_j})\\
&= \,\mathop {\min }\limits_{1 \leqslant j \leqslant k} \, \, \, \frac{1}{2}{\log ^ + }\left( {\frac{1}{{{{\mathbf{a}_j}^T}{{{\mathbf{Q}}}_j}{\mathbf{a}_j}}}} \right),
\end{aligned}
\end{equation}
where ${{\mathbf{Q}}_j}$ is defined in~(\ref{eq:Q_k}).

Note that $R^{\mathrm{(Suc)}}_1 \left( \mathbf{G} \right) \geq \cdots  \geq R^{\mathrm{(Suc)}}_L \left( \mathbf{G} \right) $. Hence, the sum rate of recovering all  messages from these $L$ equations is equal to 
\begin{equation} \label{eq:Rsum-Suc}
R^{\mathrm{(Suc)}}_{\text{sum}}=L \times R^{\mathrm{(Suc)}}_L \left( \mathbf{G} \right).
\end{equation}

Lemma~\ref{lem:Suc-rate} shows that Suc-CM scheme outperforms the Ext-CM, in terms of sum rate, for any users to relays channel distributions.

\begin{lemma} \label{lem:Suc-rate}
For any given channel matrix $\mathbf{G}$, defined in~(\ref{eq:G}), Suc-CM scheme leads to a higher or equal sum rate, compared to Ext-CM scheme. 
\end{lemma}
\begin{IEEEproof}
From~(\ref{eq:R_G_b_a_sic}), the computation rate of the optimum ECV at step $K$, i.e. ${{\mathbf{a}}_k}$, for the Suc-CM scheme, can be written as
\begin{equation}
R_k^{{\mathrm{(Suc)}}} ({\mathbf{G}})=\max_{\substack{{\mathbf{a}} \in {\mathbb{Z}^L},{\mathbf{a}} \ne {\mathbf{0}}\\
{\mathbf{a}} \perp\!\!\!\perp \left\{ {{{\mathbf{a}}_1}, \ldots ,{{\mathbf{a}}_{k - 1}}} \right\}\\
{\mathbf{b}_k} \in {\mathbb{R}^N}, {\pmb{\beta}_k} \in {\mathbb{R}^{k-1}}  }} R({\mathbf{G}},{{\mathbf{b}}_k},\,\,{{\pmb{\beta }}_k},{\mathbf{a}}).
\end{equation}
By setting ${\pmb{\beta}_k}={\mathbf{0}}$, we can write
\begin{equation}
\begin{aligned}
R_k^{{\mathrm{(Suc)}}} ({\mathbf{G}})&\geq\max_{\substack{{\mathbf{a}} \in {\mathbb{Z}^L},{\mathbf{a}} \ne {\mathbf{0}}\\
{\mathbf{a}} \perp\!\!\!\perp \left\{ {{{\mathbf{a}}_1}, \ldots ,{{\mathbf{a}}_{k - 1}}} \right\}\\
{\mathbf{b}_k} \in {\mathbb{R}^N},   }} R({\mathbf{G}},{{\mathbf{b}}_k},\,\,{{\mathbf{0}}},{\mathbf{a}})   
&=R_k^{{\mathrm{(Ext)}}} ,
\end{aligned}
\end{equation}
where the last equality follows from~(\ref{eq:R_G_b_a}). This yields the result.
\end{IEEEproof}

\subsection{Solving the Optimization Problem of Suc-CM} \label{subsec:Suc-opt}
Since the matrix $\mathbf{Q}_k$ in~(\ref{eq:ak_sic}) is not positive definite, the optimization problem of the Suc-CM scheme cannot be solved 
via standard methods, e.g. the ones proposed in~\cite{Fin}. 
We propose a method to convert this optimization problem to an uncomplicated one that includes a positive definite matrix, similar to that of the Std-CM scheme in~(\ref{eq:a1}). Also, for the readers assistant, some required basic concepts of the lattices and preliminary definitions are provided in Appendix~\ref{ap:definitions}. 

The optimization problem in~(\ref{eq:ak_sic}) can be rewritten as
\begin{equation} \label{eq:min_sic}
\min_{\substack{{\mathbf{d}} = {{\mathbf{F}}_{k - 1}}{\mathbf{a}},\,{\mathbf{a}} \in {\mathbb{Z}^L},{\mathbf{a}} \ne {\mathbf{0}}\\
{\mathbf{a}} \perp\!\!\!\perp \left\{ {{{\mathbf{a}}_1}, \ldots ,{{\mathbf{a}}_{k - 1}}} \right\}}} {{{\mathbf{d}}^{{T}}}{\mathbf{C}_k \mathbf{d}}},
\end{equation} 
where
\begin{equation} \label{eq:c_k}
{{\mathbf{C}}_k} \triangleq {\left( {{\mathbf{I}_L} + {{{\mathbf{\tilde G}}}_k}{\mathbf{\tilde G}}_k^T} \right)^{ - 1}}.
\end{equation}
The search set of~(\ref{eq:min_sic}) is
\begin{equation} \label{eq:S}
\begin{aligned}
S &= \left\{ {{\mathbf{d}}\left| {{\mathbf{d}} = {{\mathbf{F}}_{k - 1}}{\mathbf{a}},\,{\mathbf{a}} \in {\mathbb{Z}^L},{\mathbf{a}} \ne {\mathbf{0}},{\mathbf{a}} \perp\!\!\!\perp \left\{ {{{\mathbf{a}}_1}, \ldots ,{{\mathbf{a}}_{k - 1}}} \right\}} \right.} \right\} \\
& = \left\{ {{\mathbf{d}}\left| {{\mathbf{d}} = {{\mathbf{F}}_{k - 1}}{\mathbf{a}},\,{\mathbf{a}} \in {\mathbb{Z}^L},{\mathbf{d}} \ne {\mathbf{0}}} \right.} \right\}.
\end{aligned}
\end{equation}
The second equality in~(\ref{eq:S}) follows from the fact that vector $\mathbf{a}$ is linearly independent of the set $\left\{ {{{\mathbf{a}}_1}, \ldots ,{{\mathbf{a}}_{k - 1}}} \right\}$ if and only if its projection onto orthogonal complement of $\mathrm{span}\left( \mathbf{a}_1,\ldots,\mathbf{a}_{k-1} \right)$, i.e. ${{{\mathbf{F}}_{k - 1}}{\mathbf{a}}}$, is nonzero.

\begin{lemma} \label{lem:L_p}
The set ${\Lambda _P} = S \cup \left\{ {\mathbf{0}} \right\}$, which is the projection of  all $\mathbb{Z}^L$ points onto the orthogonal complement of $\mathrm{span}\left( \mathbf{a}_1,\ldots,\mathbf{a}_{k-1} \right)$, is a lattice.
\end{lemma}
\begin{IEEEproof}
We have
\begin{equation} \label{eq:L_p}
{\Lambda _P} = S \cup \left\{ {\mathbf{0}} \right\} = \left\{ {{\mathbf{d}}\left| {{\mathbf{d}} = {{\mathbf{F}}_{k - 1}}{\mathbf{a}},\,{\mathbf{a}} \in {\mathbb{Z}^L}} \right.} \right\}.
\end{equation}
Hence, for every $\mathbf{d}_1, \mathbf{d}_2 \in \Lambda_P$, there exist vectors $ \mathbf{a}_1 ,\mathbf{a}_2  \in {\mathbb{Z}^L}$ such that ${\mathbf{d}_1} = {{\mathbf{F}}_{k - 1}}{\mathbf{a}_1} $ and ${\mathbf{d}_2} = {{\mathbf{F}}_{k - 1}}{\mathbf{a}_2} $. Therefore, we can write $\mathbf{d}_1 \pm \mathbf{d}_2={\mathbf{F}}_{k - 1}\left( \mathbf{a}_1 \pm \mathbf{a}_2 \right) \in {\Lambda _P}$.
Thus, based on Definition~\ref{def:lattice} in Appendix~\ref{ap:definitions}, $\Lambda _P$ is a lattice.
\end{IEEEproof}

Definition~\ref{def:lattice} and~(\ref{eq:L_p}) yield that ${\mathbf{F}}_{k - 1}$ is a generator matrix for $\Lambda _P$. From~(\ref{eq:F_k-1}), the rank of ${\mathbf{F}}_{k - 1}$ is $L-k+1$. Hence, the rank of lattice $\Lambda _P$ is $L-k+1$. We are interested in finding a standard generator matrix for $\Lambda _P$. By applying a series of unimodular column operations on matrix ${\mathbf{F}}_{k - 1}$, we simply find an $L \times L$ matrix $B$ with $k-1$ zero columns and $L-k+1$ nonzero linearly independent  columns as
\begin{equation} \label{eq:B}
{\mathbf{B}} = {{\mathbf{F}}_{k - 1}}{\mathbf{U}} = \left( {{{{\mathbf{\tilde P}}}_{k - 1}}\left| {\mathbf{0}} \right.} \right),
\end{equation}
where the transformation matrix $\mathbf{U}$ is a unimodular matrix corresponding to the series of unimodular column operations applied on matrix ${\mathbf{F}}_{k - 1}$. From Remark~\ref{Re:genMat}, $\mathbf{B}$ is a generator matrix for $\Lambda _P$. As a result, from~(\ref{eq:lattice}) and by removing zero columns of B, the matrix ${{{{\mathbf{\tilde P}}}_{k - 1}}}$ is also a generator matrix for $\Lambda _P$. Since columns of ${{{{\mathbf{\tilde P}}}_{k - 1}}}$ are independent, it is a standard generator matrix for $\Lambda _P$. From~(\ref{eq:B}), we write
\begin{equation} \label{eq:Ptilde_k-1}
{{{\mathbf{\tilde P}}}_{k - 1}} = {{\mathbf{F}}_{k - 1}}{{\mathbf{U}}_{k - 1}},
\end{equation}
where ${\mathbf{U}}_{k - 1}$ is the matrix consisting of the first $L-k+1$ columns of $\mathbf{U}$.
Note that selecting $L-k+1$ independent columns of ${{\mathbf{F}}_{k - 1}}$ does not necessarily give the generator matrix of the lattice $\Lambda _P$ (see~\cite[Ch. 6]{Murray}).

\begin{remark}
The matrix $\mathbf{B}$ in~(\ref{eq:B}) is not unique. A special form of $\mathbf{B}$, called Hermite Normal Form (HNF), can be calculated from the integer matrix  ${{\mathbf{F'}}_{k - 1}} = \det \left( {{\mathbf{A}}_{k - 1}^T{{\mathbf{A}}_{k - 1}}} \right){{\mathbf{F}}_{k - 1}}$. It can be proved that HNF of an integer matrix always exists and is unique~\cite[Sec. 2.4.2]{Cohen}. A pseudo code for finding HNF of a matrix and its corresponding transformation matrix can be found in~\cite[Sec. 2.4.2]{Cohen}.  
\end{remark}

Since ${{{{\mathbf{\tilde P}}}_{k - 1}}}$ is a basis for lattice $\Lambda _P$, we represent the lattice as
\begin{equation} \label{eq:L_p2}
{\Lambda _P} = \left\{ {{\mathbf{d}}\left| {{\mathbf{d}} = {{{\mathbf{\tilde P}}}_{k - 1}}{\mathbf{w}},\,\,\,{\mathbf{w}} \in {\mathbb{Z}^{L - k + 1}}} \right.} \right\}.
\end{equation}
The search set $S$ is expressed as
\begin{equation} \label{eq:S2}
S = \left\{ {{\mathbf{d}}\left| {{\mathbf{d}} = {{{\mathbf{\tilde P}}}_{k - 1}}{\mathbf{w}},\,\,\,{\mathbf{w}} \in {\mathbb{Z}^{L - k + 1}},\,{\mathbf{w}} \ne {\mathbf{0}}} \right.} \right\}.
\end{equation}

To convert the optimization problem of Suc-CM scheme to a simpler form, as in Std-CM scheme, we propose the following theorem.
It is worth noting that Theorem~\ref{th:Opt_sic} decreases the complexity of finding ECV ${{\mathbf{a}}_k}$ significantly, since 1) the optimization includes a positive definite matrix ${{\mathbf{\tilde Q}}_k}$ rather than the positive semi-definite matrix $\mathbf{Q}_k$  in~(\ref{eq:ak_sic}), and hence  numerical methods such as~\cite{Fin} can be exploited. Moreover, 2) the search space of the optimization reduces from ${\mathbb{Z}^{L}}$ to ${\mathbb{Z}^{L - k + 1}}$. This is achieved due to mapping of $\mathbb{Z}^L$ points to a lattice of dimension  $L-k+1$.
Finally, 3) there is no need to check the linear independency of ${{\mathbf{a}}_k}$ from previous ECVs $\left\{ {{{\mathbf{a}}_1}, \ldots ,{{\mathbf{a}}_{k - 1}}} \right\}$, in the iterations of exploited numerical methods.

\begin{theorem} \label{th:Opt_sic}
The optimum ECV $\mathbf{a}_k$ of step $k$ , i.e. the solution of Suc-CM optimization problem~(\ref{eq:ak_sic}), can be equivalently calculated from the following optimization problem:
\begin{equation} \label{eq:Opt_sic}
\begin{aligned}
\mathbf{w}_k &= \arg \min_{\substack{{\mathbf{w}} \in {\mathbb{Z}^{L-k+1}},\\{\mathbf{w}} \ne {\mathbf{0}}}} {{{\mathbf{w}}^{{T}}}{\mathbf{\tilde Q}_k \mathbf{w}}},
\\{{\mathbf{a}}_k} &= {{\mathbf{U}}_{k - 1}}{{\mathbf{w}}_k},
\end{aligned}
\end{equation}
where
\begin{equation} \label{eq:Qtilde_k}
{{\mathbf{\tilde Q}}_k} = {\mathbf{\tilde P}}_{k - 1}^T{{\mathbf{C}}_k}{{\mathbf{\tilde P}}_{k - 1}}
\end{equation}
is a positive definite matrix, ${{\mathbf{C}}_k}$ is defined in~(\ref{eq:c_k}), and ${{\mathbf{\tilde P}}_{k - 1}}$ and ${{\mathbf{U}}_{k - 1}}$ are found from~(\ref{eq:Ptilde_k-1}).
\end{theorem}

\begin{IEEEproof}
See Appendix~\ref{ap:Opt_sic_proof}.
\end{IEEEproof}

From~(\ref{eq:R_G_a_sic}) and~(\ref{eq:Opt_sic}), 
the computation rate corresponding to a vector ${\mathbf{w}} \in {\mathbb{Z}^{L - k + 1}}$, in the search set of~(\ref{eq:Opt_sic}), is given by 
\begin{equation}
R \left(\mathbf{w}\right) =  \frac{1}{2}{\log ^ + }\left( {\frac{1}{{{{\mathbf{w}}^T}{{{\mathbf{\tilde Q}}}_k}{\mathbf{w}}}}} \right),
\end{equation}
where ${\lambda _{\min }}( {{{{\mathbf{\tilde Q}}}_k}} )$ is the minimum eigenvalue of the matrix ${{{\mathbf{\tilde Q}}}_k}$. 
From Courant-Fischer theorem~\cite[Ch. 7]{Meyer}, it follows that
${\lambda _{\min }}( {{{{\mathbf{\tilde Q}}}_k}} ){\left\| {\mathbf{w}} \right\|^2} \leqslant {{\mathbf{w}}^T}{{\mathbf{\tilde Q}}_k}{\mathbf{w}}$.
Then, we can write
\begin{equation}
\begin{aligned}
R \left(\mathbf{w}\right) &= \frac{1}{2}{\log ^ + }\left( {\frac{1}{{{{\mathbf{w}}^T}{{{\mathbf{\tilde Q}}}_k}{\mathbf{w}}}}} \right) \\
 &\leqslant \frac{1}{2}{\log ^ + }\left( {\frac{1}{{{\lambda _{\min }}\left( {{{{\mathbf{\tilde Q}}}_k}} \right){{\left\| {\mathbf{w}} \right\|}^2}}}} \right).
\end{aligned}
\end{equation}
Hence, every vector  ${\mathbf{w}}$ with ${\lambda _{\min }}( {{{{\mathbf{\tilde Q}}}_k}} ){\left\| \mathbf{w} \right\|^2} \leqslant 1$ results in zero computation rate.
As result, the search set of the optimization in~(\ref{eq:Opt_sic}) can be limited to the vectors $\mathbf{w}$ that satisfy
\begin{equation} \label{eq:w_lim2}
{\left\| {\mathbf{w}} \right\|^2} < \frac{1}{{{\lambda _{\min }}\left( {{{{\mathbf{\tilde Q}}}_k}} \right)}}.
\end{equation}
Note that~(\ref{eq:w_lim2}) 
is also applicable to 
optimization problems in~(\ref{eq:a1}) and~(\ref{eq:ak}), by replacing ${{{{\mathbf{\tilde Q}}}_k}}$ with $\mathbf{Q}$.
The optimization in~(\ref{eq:Opt_sic}) has the same form as~(\ref{eq:a1}) and, similarly, can be solved by the algorithms suggested in~\cite{Fin, Lili, P09 , P10, Algebraic,Soussi,P07}. However,~(\ref{eq:w_lim2}) can be used to determine some required parameters for these algorithms, like an initial radius $r$ to start the search  for the algorithm proposed in~\cite{Fin}.

\section{Performance Analysis  of the Proposed Methods} \label{sec:perf}
To evaluate the performance of our proposed CMF methods, we provide the outage performance and  diversity analysis of the extended CMF and successive methods over the multi-user multi-relay networks. We consider the cases of ideal and non-ideal R-D channels, respectively, in the following subsections.

Let us define ${R_{\text{rel}}}({{\mathbf{G}}^{m}})$ as the  achievable sum rate of the $m$-th relay, where ${{\mathbf{G}}^{m}}$ is the corresponding relay channel matrix, defined in~(\ref{eq:G}).
Note that this sum rate is equal to~(\ref{eq:Rsum-Ext}) and~(\ref{eq:Rsum-Suc}), for the cases with Ext-CM and Suc-CM,
respectively. Hence, the following analysis covers both extended and successive CMF methods.

The outage probability of the relay $m$, for its sum rate, is defined as 
\begin{equation} \label{Pout_relay}
P_{\text{relay},m}^{\text{out}} \triangleq \Pr \left\{ {{R_{\text{rel}}}({{\mathbf{G}}^{m}}) < {R_{\text{t}}}} \right\},
\end{equation}
where $R_{\text{t}}$ is the target sum rate. For a system with outage probability $P_{\text{out}}$, the diversity order of the system is defined as
\begin{equation} \label{eq:d-lim}
d \triangleq  - \mathop {\lim }\limits_{\gamma  \to \infty } \frac{{\log {P_\text{out}}}}{{\log \gamma }},
\end{equation}
where $\gamma$ is the average SNR of the channels~\cite{Tse}. For simplicity of presentation, (\ref{eq:d-lim}) can be written in the alternative form
\begin{equation} \label{eq:d-power}
{P_{\text{out}}} \doteq {\gamma ^d},
\end{equation}
at high SNRs, where the symbol $\doteq$ indicates the asymptotic  equality for $\gamma  \to \infty $. Moreover, to find the diversity order of the system, we consider the same SNR $\gamma$ for all channels.

\begin{remark} \label{Re:d-full}
The diversity order of a system with $L$ independently transmitted streams and $N \geq L$ receive antennas, over real Gaussian channels, is at most equal to $\frac{N}{2}$. Note that the factor $\frac{1}{2}$ is removed for complex channels. Moreover, cooperation among transmitters, e.g. joint space-time coding, can increase the diversity limit significantly\cite{Tse,IFLR}.
\end{remark}
\begin{lemma} \label{lem:d-SucCM}
For an $N$-antenna relay ($N \geq L$) and $L$ users, over real user-to-relay (U-R) Gaussian channels, the Ext-CM and Suc-CM schemes achieve the full diversity order $\frac{N}{2}$.
\end{lemma}
\begin{IEEEproof}
The diversity order of the IFLR scheme, for the conditions stated in the lemma, is proved to be $\frac{N}{2}$ \cite{IFLR}. Since the users transmit independently, this is the highest possible diversity order of the system. Thus, from Theorem~\ref{th:IFLR} and Lemma~\ref{lem:Suc-rate}, it follows that the diversity orders of Ext-CM and Suc-CM schemes are  $\frac{N}{2}$ as well.
\end{IEEEproof}

Note that, from Section~\ref{subsec:channels}, the diversity order $N$ can be achieved if the complex case of CMF is used, or both real and imaginary parts of the received signals are employed by relays.

\subsection{Ideal R-D channels}
Now, consider $M$ relays in the network. The network employs one of the extended CMF or the successive CMF methods. The U-R channels are real Gaussian, and the R-D channels are considered to be ideal (see Section~\ref{sec:model}).
Since in Sel-FW the best relay sends its equations to the destination, the outage probability of the system is found as
\begin{equation} \label{eq:Pout_sys_Ideal}
\begin{aligned}
P_{\text{sys,Ideal}}^{\text{out},M} &= \Pr \left\{ \max_{\substack{  1 \leq m \leq M }} {{R_{\text{rel}}}({{\mathbf{G}}^{m}})} < {R_{\text{t}}} \right\} \\
&= \Pr \left\{  {{R_{\text{rel}}}({{\mathbf{G}}^{1}}) < {R_{\text{t}}}}, \ldots,  {{R_{\text{rel}}}({{\mathbf{G}}^{M}}) < {R_{\text{t}}}}  \right\} \\
 &\mathop  = \limits^{(a)}   \prod_{m = 1}^M {\Pr \left\{ {{R_{\text{rel}}}({{\mathbf{G}}^{m}}) < {R_{\text{t}}}} \right\}} \\
 &= \prod\nolimits_{m = 1}^M {P_{\text{relay},m}^{\text{out}}}\, ,
\end{aligned}
\end{equation}
where $(a)$ holds since the relay channel matrices are statistically independent, and the relays select their ECVs statistically independent of each other. 

To find the diversity order of the network, we propose the following theorem.
\begin{theorem} \label{th:d-Ideal}
The extended CMF and successive CMF methods over the $L$-user $M$-relay network with real Gaussian U-R channels and ideal R-D channels, achieves the  diversity order 
\begin{equation} \label{eq:d-Ideal}
{d_{\text{sys,Ideal}}} = \frac{{MN}}{2},
\end{equation}
where $N$, $N \geq L$, is the number of exploited antennas in each relay.  
\end{theorem}
\begin{IEEEproof}
See Appendix~\ref{ap:d-Ideal}.
\end{IEEEproof}
From Remark~\ref{Re:d-full}, since the total number of relay antennas in the network is $MN$, both   extended CMF and successive CMF methods achieve the full diversity of the real Gaussian network, i.e.  $\frac{MN}{2}$.

\subsection{Non-Ideal R-D channels}
To evaluate the effect of non-ideal R-D channels,  we consider i.i.d. Nakagami($q$) distributions with unit variance for channels coefficients $f_m, m=1,\ldots,M$. From Section~\ref{subsec:channels}, $f_m, m=1,\ldots,M$ are fixed during each transmission frame, and are independent of the ones in other transmission frames. The transmission rates over the R-D channels are ${r_m} = \frac{1}{2}\log \left( {1 + \gamma {{\left| {{f_m}} \right|}^2}} \right), m=1, \ldots, M$, where $\gamma$ is the average SNR of the channels. The selected relay uses its R-D channel $L$ times to send its $L$ selected equations to the destination. Thus the outage probability of each R-D channel is  calculated as
\begin{equation} \label{Pout-fm}
\begin{aligned}
P_{\text{R-D}}^{\text{out}} &\triangleq \Pr \left\{ {L{r_m} < {R_{\text{t}}}} \right\} \\
&= \Pr \left\{ {\frac{L}{2}\log \left( {1 + \gamma{{\left| {{f_m}} \right|}^2}} \right) < {R_{\text{t}}}} \right\}\\
 &= \Pr \left\{ {{{\left| {{f_m}} \right|}^2} < \frac{{{2^{\frac{{2{R_{\text{t}}}}}{L}}} - 1}}{\gamma }} \right\}\\
  &= {F_{{\chi ^2}}}\left( {2q,\frac{{{2^{\frac{{2{R_{\text{t}}}}}{L}}} - 1}}{\gamma }} \right),
 \end{aligned}
\end{equation} 
where ${F_{{\chi ^2}}}\left( {2q,x} \right) $ is the cumulative distribution function (CDF) of chi-square distribution with $2q$ degrees of freedom. 
The last equality  follows from the fact that ${{\left| {{f_m}} \right|}^2}$ is a Chi-Square random variable with $2q$ degrees of  freedom~\cite{Karbalayghareh}.

To find the diversity order, we consider two different cases, as follows.

\subsubsection{No D-R Feedback}
 If the R-D channel of the selected relay is in outage, the equations are not received by the destination, and hence, a system outage event occurs.
Since, we have assumed only local CSI for the relays, and if there is no destination-to-relays (D-R) feedback, the relays are not aware of the R-D channels states. Thus, for the case of 
no D-R feedback, the outage probability of the system is 
\begin{equation} \label{eq:Pout-sys-NF}
P_{\text{sys,NF}}^{\text{out}} = P_{\text{R-D}}^{\text{out}} + \left( {1 - P_{\text{R-D}}^{\text{out}}} \right).P_{\text{sys,Ideal}}^{\text{out},M},
\end{equation}
where $P_{\text{R-D}}^{\text{out}}$ and $P_{\text{sys,Ideal}}^{\text{out},M}$ are given in~(\ref{Pout-fm}) and~(\ref{eq:Pout_sys_Ideal}), respectively. The related diversity order is determined by the  following theorem.

\begin{theorem} \label{th:d-NF}
Without D-R feedback, the extended CMF and successive CMF methods, over the $L$-user $M$-relay network with real Gaussian U-R channels and Nakagami($q$) R-D channels, achieve diversity order 
\begin{equation} \label{eq:d-NF}
{d_{\text{sys,NF}}} = \min \left( {q,\frac{{MN}}{2}} \right),
\end{equation}
where $N$, $N \geq L$, is the number of exploited antennas in each relay.  
\end{theorem}
\begin{IEEEproof}
See Appendix~\ref{ap:d-NF}.
\end{IEEEproof}

\subsubsection{With D-R Feedback}
To improve the performance of system in the case of non-ideal R-D channels, we modify the Sel-FW strategy to use negligible D-R feedback. In modified Sel-FW strategy, it is assumed that the destination have local CSI, i.e. $f_m, m=1, \ldots, M$, and hence, knows which R-D channels are in outage. The destination informs the relays of their corresponding R-D channel state (i.e. outage or good). Hence, only the relays with good channel state, find their ECVs and participate in the best relay selection process (e.g. set their timers, see Section~\ref{subsec:Sel-FW}). Note that if a subset of $m$ out of $M$ relays have good R-D channel states, the system outage performance is the same as a system with $m$ relays and ideal R-D channels. Hence, using the law of total probability, the outage probability of the non-ideal R-D channel case with D-R feedback is expressed as
\begin{equation} \label{eq:Pout-WF0}
P_{\text{sys,WF}}^{\text{out}} = \sum\limits_{m = 0}^M {{p_m}.P_{\text{sys,Ideal}}^{\text{out},m}},
\end{equation}
where $P_{\text{sys,Ideal}}^{\text{out},m}$ is the outage probability of a system with $m$ relays and ideal R-D channels, which is found from~(\ref{eq:Pout_sys_Ideal}), and we have $P_{\text{sys,Ideal}}^{\text{out},0}=1$. Also, $p_m$ is the probability of the event that exactly $m$ out of $M$ relays have good R-D channel states. Since, the channel coefficients $f_m, m=1,\ldots,M$ are independent, the probability $p_m$ can be written as
\begin{equation} \label{eq:p_m1}
{p_m} = {M \choose m} {\left( {1 - P_{\text{R-D}}^{\text{out}}} \right)^m}{\left( {P_{\text{R-D}}^{\text{out}}} \right)^{M - m}},
\end{equation} 
where $M \choose m$ is the "n choose k" operator. Substituting~(\ref{eq:p_m1}) in~(\ref{eq:Pout-WF0}) yields
\begin{equation} \label{eq:Pout-WF1}
P_{\text{sys,WF}}^{\text{out}} = \sum\limits_{m = 0}^M {{M \choose m}{{\left( {1 - P_{\text{R-D}}^{\text{out}}} \right)}^m}{{\left( {P_{\text{R-D}}^{\text{out}}} \right)}^{M - m}}P_{\text{sys,Ideal}}^{\text{out},m}}.
\end{equation}
The following theorem gives the diversity order achieved by exploiting modified Sel-FW strategy in the case of non-ideal R-D channel.
\begin{theorem} \label{th:d-WF}
With D-R feedback, real Gaussian U-R channels, and Nakagami($q$) R-D channels, the extended CMF and successive CMF methods (employing modified Sel-FW strategy over the $L$-user $M$-relay network) achieve the diversity order
\begin{equation} \label{eq:d-WF}
{d_{\text{sys,WF}}} = M \times \min \left( {q,\frac{N}{2}} \right),
\end{equation}
where $N$, $N \geq L$, is the number of exploited antennas in each relay.  
\end{theorem}
\begin{IEEEproof}
See Appendix~\ref{ap:d-WF}.
\end{IEEEproof}

\begin{remark} \label{Re:Mojahedian}
From~(\ref{eq:d-NF}) and~(\ref{eq:d-WF}), we always have ${d_{\text{sys,WF}}} \geq {d_{\text{sys,NF}}}$. However, D-R feedback helps, i.e. increases the diversity order, for $q < \frac{MN}{2}$. This means that when the long-term R-D channels conditions are better than a certain quality, specifically for $q \geq \frac{MN}{2}$, using D-R feedback provides no diversity gains. This is due to the fact that for $q \geq \frac{MN}{2}$ the system achieves full diversity, i.e. $\frac{MN}{2}$. 
\end{remark}

It is worth noting that in the case of $q < \frac{MN}{2}$, the use of D-R feedback can be replaced with employing larger number of antennas for R-D channels to get $q=\mu N_t N_r \geq \frac{MN}{2}$. Hence, the system can achieve full diversity without using D-R feedback.
Moreover, it is interesting that when the diversity is limited by the quality of the R-D channels , i.e. we have $q \leq \frac{MN}{2}$ for no R-D feedback case and $q \leq \frac{N}{2}$ for the cases with R-D feedback, employing larger number of relays $M$ or relay antennas $N$ provides only SNR gains.

\section{Numerical Results} \label{sec:simul}
To evaluate the performance of the proposed methods, we provide computer simulations for different scenarios. We compare our proposed extended and successive CMF methods with the original CMF method~\cite{BN1}. In simulations, equal power $P$, for transmitting nodes, and unit variance for channel noises are considered, i.e. we have $P_l=P_R=P, l=1,\ldots,L$, and $\sigma_{nm}^2={\sigma'}_m^2=1, m=1,\ldots,M,\,n=1,\ldots,N$. The zero-mean Gaussian distribution with unit variance is considered for the coefficients of the users to relays channels (see Section~\ref{subsec:channels}). Since unit variance is assumed for the channel gains, the average SNR of the channels is equal to $P$. 

In simulations, the performance is measured in terms of average sum rate of the users and overall outage probability of the system. The unit of average sum rate is bits per transmission frame (consisting of $L+1$ time slots). Note that the original CMF method uses a total of $M+1$ time slots, while the successive and extended CMF methods require $L+1$ time slots, for each transmission frame. Thus, for a fair comparison, the sum rate of the original CMF method is multiplied by $\tfrac{{\left( {L + 1} \right)}}{{\left( {M + 1} \right)}}$ in simulations. 
To find the ECVs in the original, extended, and successive CMF methods, we have solved the optimization problems~(\ref{eq:a1}),~(\ref{eq:ak}), and~(\ref{eq:Opt_sic}), respectively, using the approach given in~\cite{Fin}.
Note that since Ext-CM scheme has the same performance as IFLR scheme (see Theorem~\ref{th:IFLR}), the curves corresponding to Ext-CM scheme in figures are also true for IFLR scheme.

\begin{figure}[tb]
\renewcommand{\figurename}{Fig.}
\centering
\includegraphics[trim = 8mm 2mm 15mm 9mm, clip, width =\Wf \columnwidth]{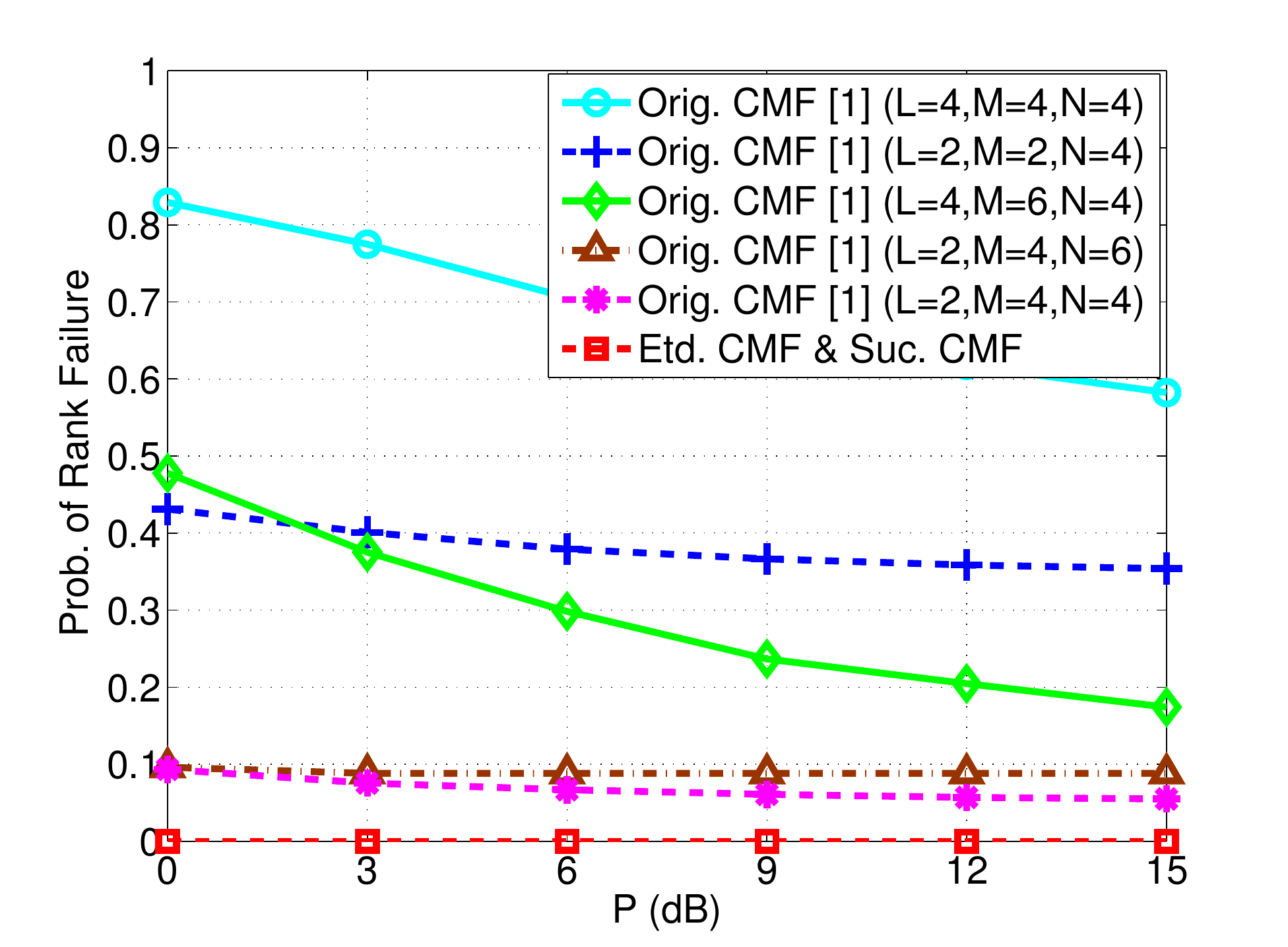}
\caption{\small Probability of rank failure at the destination versus average SNR, (ideal R-D channels).}
\label{fig:PrRankFail_Ideal}
\end{figure}

Fig.~\ref{fig:PrRankFail_Ideal} shows the probability of rank failure at the destination for the original, extended, and successive CMF methods, versus average SNR. The R-D channels are assumed to be ideal.
The extended and successive CMF methods yield zero probability of rank failure, since in these methods, $L$ linearly independent equations are recovered by the selected relay and sent to the destination. On the other hand, in original CMF, each relay selects and sends one equation statistically independently of the other relays. Hence, the received equations by destination may be linearly independent, and rank failure may occur. This figure indicates that rank failure for the original method occurs in most cases, with high probability. 
As it is observed, by increasing the number of relays $M$, probability of rank failure decreases in the original CMF method. The reason is that the destination receives more equations, and, with a higher probability can find $L$ linearly independent equations among them. In addition, the rank failure probability increases with $L$, since the destination needs to find more linearly independent equations among its received equations. Moreover, Fig.~\ref{fig:PrRankFail_Ideal} reveals that employing  larger number of antennas at the relays ,$N$, does not necessarily decrease the rank failure probability in the original CMF method. 
The distribution and SNR of U-R channels 
affects the distribution of selected ECVs at the relays. 
As a result, the rank failure probability is dependent on the channels distribution and SNR, as observed in Fig.~\ref{fig:PrRankFail_Ideal}.
Note that, although the rank failure probability of the original CMF method decreases with SNR for the parameter settings of Fig.~~\ref{fig:PrRankFail_Ideal}, this is not true for all 
cases (see~\cite{myIET} for more details).

\begin{figure}[tb]
\renewcommand{\figurename}{Fig.}
\centering
\includegraphics[trim = 5mm 2mm 10mm 8mm, clip, width =\Wf \columnwidth]{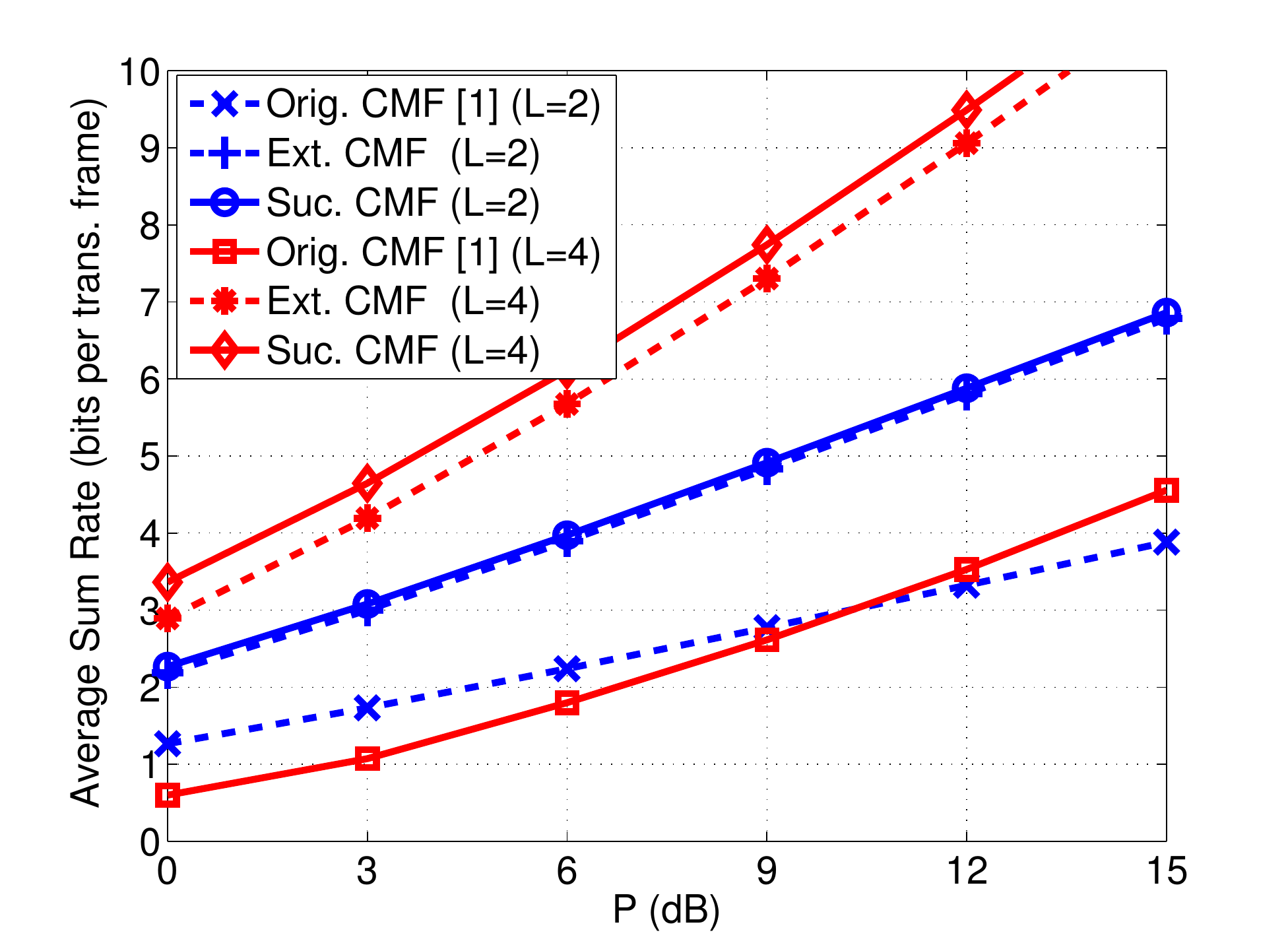}
\caption{\small Average sum rates of the original, extended, and successive CMF methods versus average SNR, for $L=2$ and $L=4$ users, (ideal R-D channels, $M=4$, $N=4$).}
\label{fig:RsAve_vs_SNR_M4_N4_L2and4_Ideal}
\end{figure}

\begin{figure}[tb]
  \renewcommand{\figurename}{Fig.}
\centering
\includegraphics[trim = 5mm 2mm 10mm 6mm, clip, width =\Wf \columnwidth]{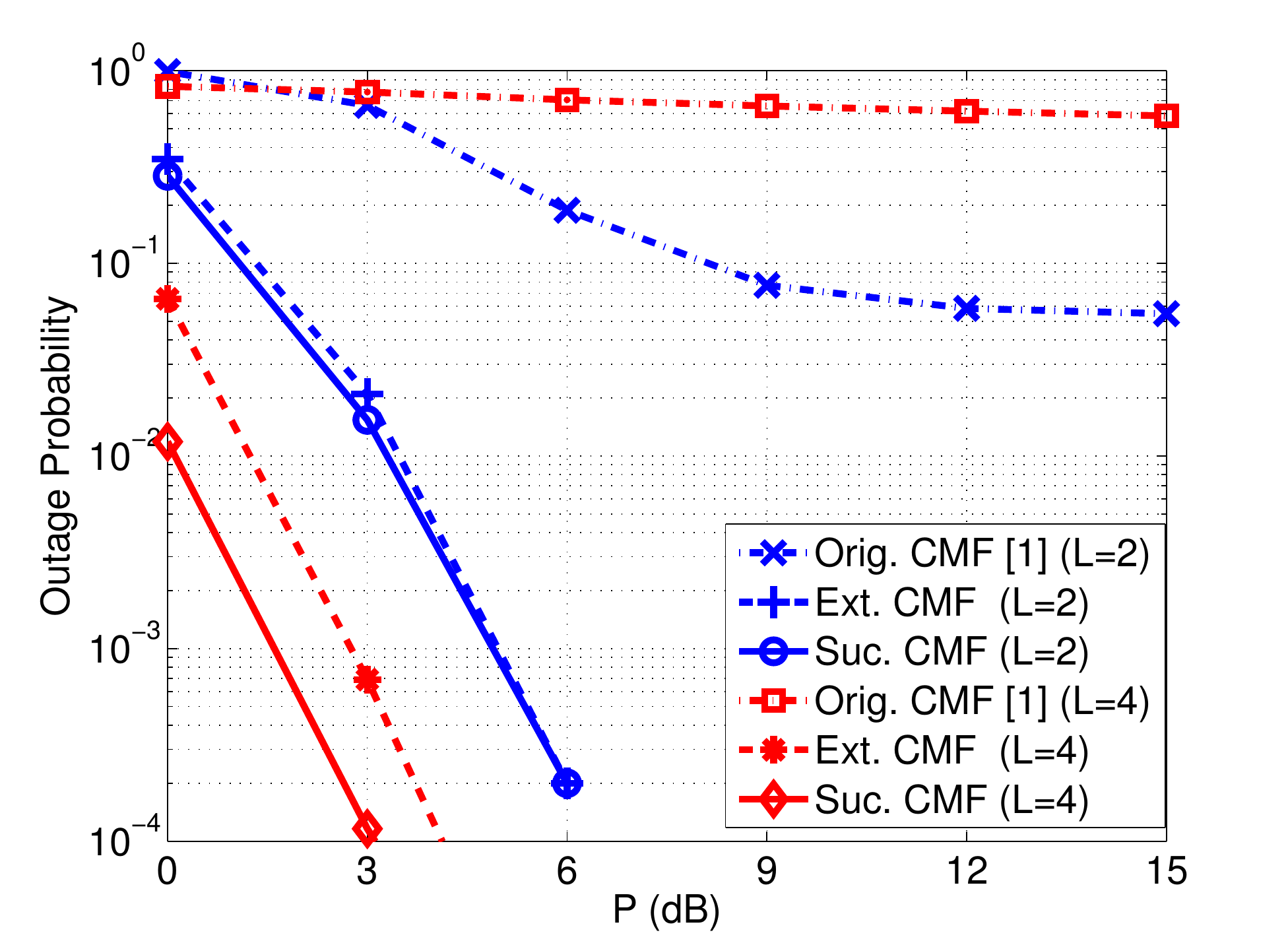}
\caption{\small Outage probabilities of the original, extended, and successive CMF methods versus average SNR, for $L=2$ and $L=4$ users, (ideal R-D channels, $R_{\text{t}}=2$, $M=4$, $N=4$).}
\label{fig:Pout_vs_SNR_M4_N4_L2and4_Ideal_Rs2}
\end{figure}

Figs.~\ref{fig:RsAve_vs_SNR_M4_N4_L2and4_Ideal} and~\ref{fig:Pout_vs_SNR_M4_N4_L2and4_Ideal_Rs2} compare the original, extended, and successive CMF methods, in terms of average sum rate and outage probability, respectively. Two cases of $L=2$ and $L=4$ users are considered. Ideal R-D channels, the target sum rate $R_{\text{t}}=2$, $M=4$, and $N=4$ are assumed. As it is observed in figures, the extended and successive CMF methods perform significantly better than the  original CMF method. The reason of poor performance of the original CMF method and its rate loss, especially at low SNRs, is the rank failure problem.
It is shown in~\cite{myIET} that the overall outage probability of the system is lower bounded by the rank failure probability. The rank failure problem decreases the diversity order (slope of the curves at high SNRs) of the system considerably. 
Moreover, as it is observed from Fig.~\ref{fig:RsAve_vs_SNR_M4_N4_L2and4_Ideal}, due to the rank failure problem, increasing the number of users does not necessarily increase the average sum rate in the original CMF method.
This is because, in general, the average sum rate of the original CMF method depends on two factors, namely, the rank failure probability and the computation rates of equations recovered by $M$ relays, both of  which depend on SNR. Specifically, the trade-off between these two factors determines the performance at different SNR regions.
Furthermore, Successive CMF method shows a better performance than the extended CMF method, especially for large values of $L$. This is due to the fact that as $L$ increases, i.e. the number of recovered equations in each relay increases, the successive CMF method gets more degrees of freedom to adjust the equation coefficients.

\begin{figure}[tb]
\renewcommand{\figurename}{Fig.}
\centering
\includegraphics[trim = 5mm 2mm 10mm 8mm, clip, width =\Wf \columnwidth]{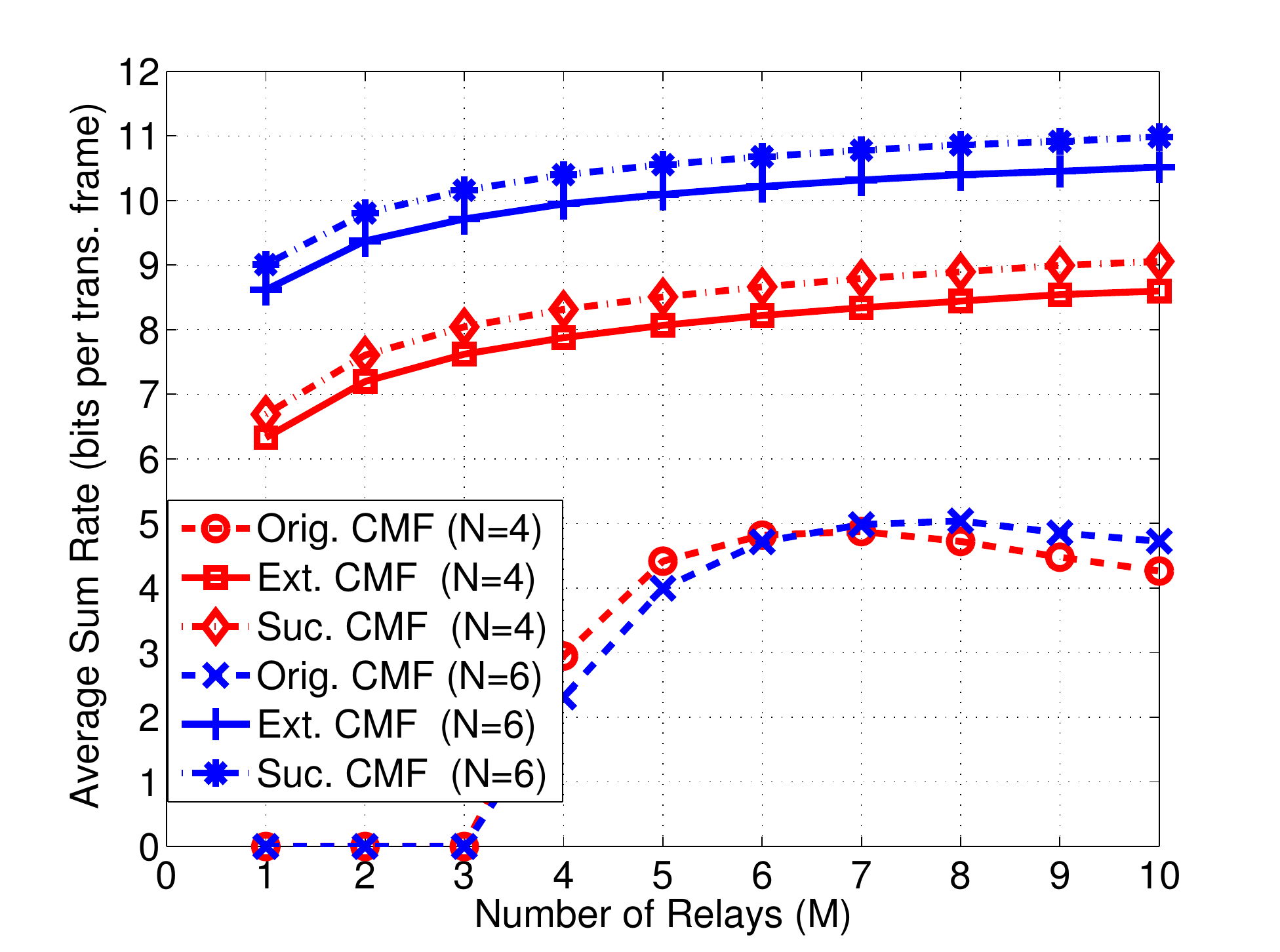}
\caption{\small Average sum rates of the original, extended, and successive CMF methods versus the number of relays, at average SNR $P=10$ dB, (ideal R-D channels, $L=4$).}
\label{fig:Fig_L4_N4and6_Ideal_RsAve_vs_M}
\end{figure}

In Fig.~\ref{fig:Fig_L4_N4and6_Ideal_RsAve_vs_M}, the effect of employing larger numbers of relays on the network performance is illustrated. The average sum rates of the original, extended, and successive CMF methods versus the number of relays, $M$ , are plotted at the average SNR $P=10$ dB.  Two cases of relays with $N=2$ and $N=4$ antennas are considered. Ideal R-D channels and $L=4$ users are assumed. Average sum rate is  a strictly increasing function of $M$ for the extended and successive CMF methods, since increasing the numbers of relays leads to higher diversity orders for the system. Moreover, utilizing larger number of antennas at each relay, improves the performance of these two methods considerably.  The reason is that by receiving more signals at each relay, the relay can combine them more efficiently and recover an equation with  higher rate. For the original CMF, using higher $M$, on the one hand, decreases the rank failure probability, and hence, provides rate gains. On the other hand, exploiting more relays increases the number of required time slots for the transmission  frame, and thus, reduces the rate. This trade off results in a optimum, in terms of sum throughput, value for $M$, e.g. $M=7$ and $M=8$ for $N=4$ and $N=6$, respectively. 
Furthermore, as observed in Fig.~\ref{fig:Fig_L4_N4and6_Ideal_RsAve_vs_M}, due to the rank failure problem, exploiting larger number of antennas does not necessarily increase the average sum rate in the original CMF method. 
Specifically, the trade-off between the rank failure probability and the computation rates of relays' equations  determines the performance for different values of $M$.
Note that since original CMF method is not applicable where $M<L$, zero rates are considered for these cases.

\begin{figure}[tb]
\renewcommand{\figurename}{Fig.}
\centering
\includegraphics[trim = 5mm 2mm 10mm 6mm, clip, width =\Wf \columnwidth]{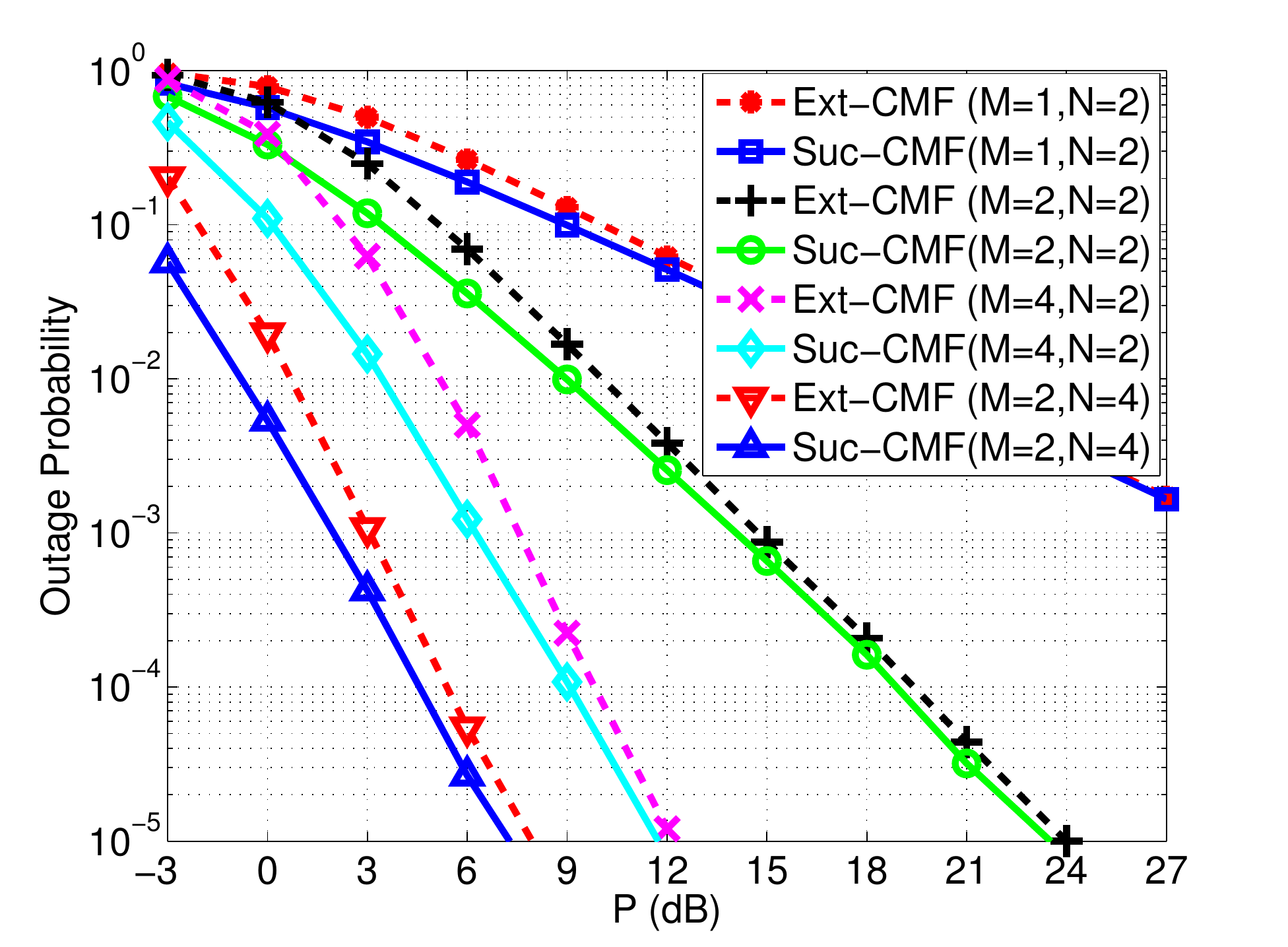}
\caption{\small Outage probabilities of the extended and successive CMF methods, versus average SNR, for  different number of relays $M$ and relay antennas $N$, (ideal R-D channels, $R_{\text{t}}=1$, $L=4$).}
\label{fig:SucCMF_N_M_Ideal}
\end{figure}

In Fig.~\ref{fig:SucCMF_N_M_Ideal}, the effect of the number of relays, $M$, and the number of each relay antennas, $N$, on the network outage probability is shown, for the extended and successive CMF methods. The outage probabilities of the both methods, versus average SNR, are plotted for $L=4$ users. $R_{\text{t}}=1$ and ideal R-D channels are assumed. As it is expected, increasing each of $M$ and $N$ improves the performance and increases the diversity order for both methods. For instance, consider the case $(M=2,N=2)$. Adding two users, or two antenna to each relay provides nearly  $8$ dB, or $12$ dB, SNR gain, respectively, at outage probability of $0.001$, for successive CMF method. As observed from this figure, for a given $M$ and $N$, both extended and successive CMF methods achieve the same order of diversity, while the latter provides an additional SNR gain. This gain increases as SNR decreases.

By comparing the cases $(M=4,N=2)$ and $(M=2,N=4)$ in Fig.~\ref{fig:SucCMF_N_M_Ideal}, it can be found that, with the considered system model of the paper, collecting the antennas in a small number of relays is more beneficial than distributing them among large number of relays. Note that the slopes of all curves for the cases  $(M=4,N=2)$ and $(M=2,N=4)$, at high SNR values, are nearly the same. This indicates that all these cases achieve the same diversity order, which is, from the figure, equal to $4$. This is in agreement with the theoretical value of diversity in~(\ref{eq:d-Ideal}), i.e. $\tfrac{MN}{2}=4$. The case with higher $N$, i.e. $(M=2,N=4)$, provides higher SNR gain, since more antennas are centralized in each relay and their received signals can be combined to increase the equation computation rates.

\begin{figure}[tb]
\renewcommand{\figurename}{Fig.}
\centering
\includegraphics[trim = 8mm 2mm 15mm 6mm, clip, width =\Wf \columnwidth]{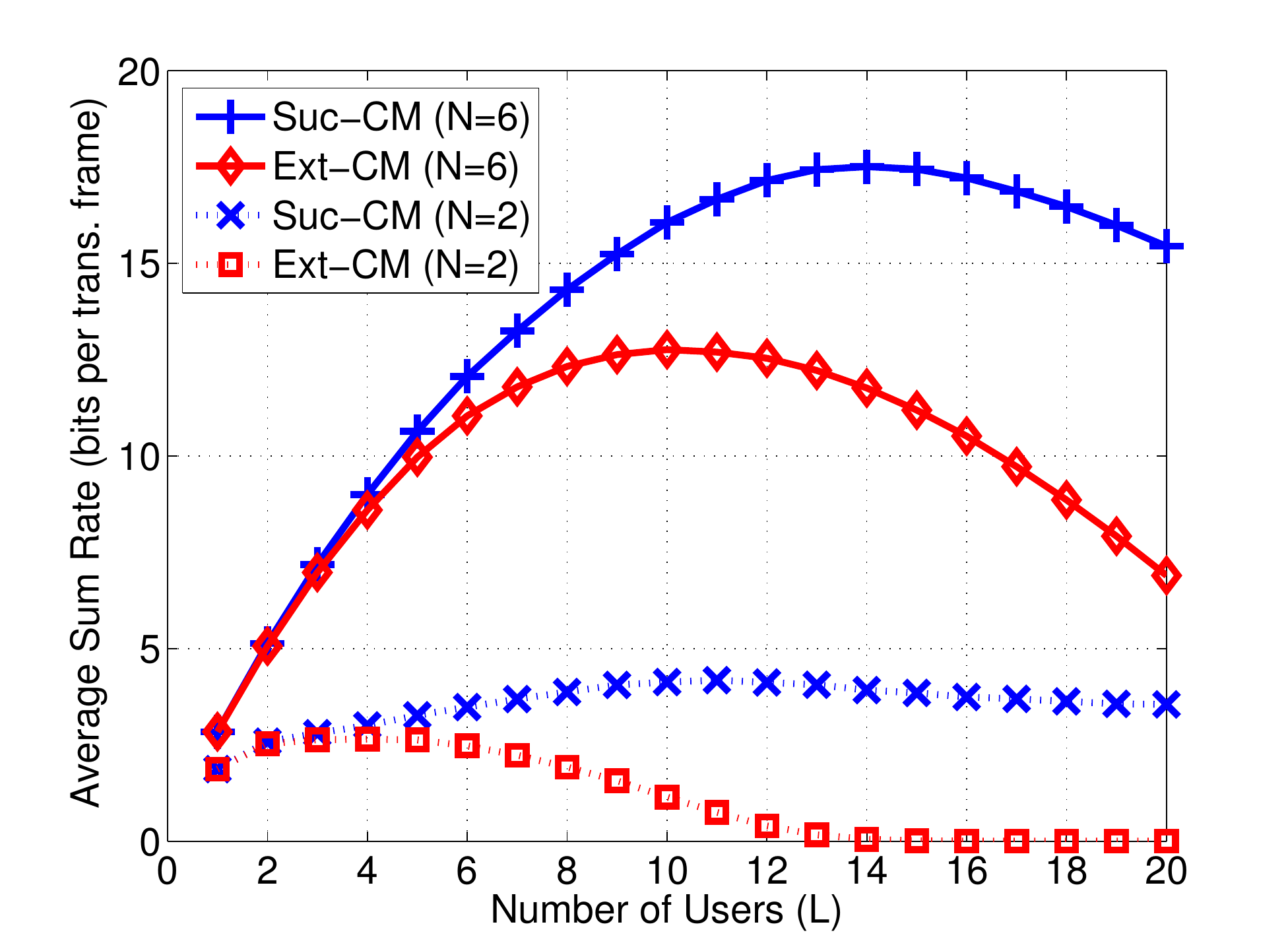}
\caption{
\small Average sum rates of the Ext-CM and Suc-CM schemes versus the number of users, at average SNR $P = 10$ dB ($M =1$). 
}
\label{fig:Rsave_vs_L}
\end{figure}

In Fig.~\ref{fig:Rsave_vs_L}, the effect of number of users, $L$, on the performance of the Ext-CM and Suc-CM schemes is demonstrated. 
The average sum rates achieved by a relay exploiting either of the schemes, versus $L$, are plotted at the average SNR $P = 10$  dB.
Two cases of~$N=2$ and~$N=6$ antennas at the relay are considered. 
As it is shown, the Suc-CM outperforms the Ext-CM scheme in terms of average sum rate,  and the performance gap increases with $L$. 
This is due to the fact that Suc-CM uses previously decoded equations to improve the computation rates of subsequent equations.
Therefore, Suc-CM scheme is more proper than Ext-CM for networks with high number of users.
Note that, although the average rate of the last equation decoded at a relay decreases with $L$ for both schemes, the  average sum rate increases for small values of $L$, as it is observed in the figure. 
This is true in general, for $L \leq N$. Hence, exploiting sufficient number of antennas (at least equal to the number of users) at relays,  the average sum rate increases with $L$. This is in agreement with Figs.~\ref{fig:RsAve_vs_SNR_M4_N4_L2and4_Ideal} and~\ref{fig:Pout_vs_SNR_M4_N4_L2and4_Ideal_Rs2}, in which $L \leq N$.

\begin{figure}[tb]
\renewcommand{\figurename}{Fig.}
\centering
\includegraphics[trim = 6mm 2mm 15mm 6mm, clip, width =\Wf \columnwidth]{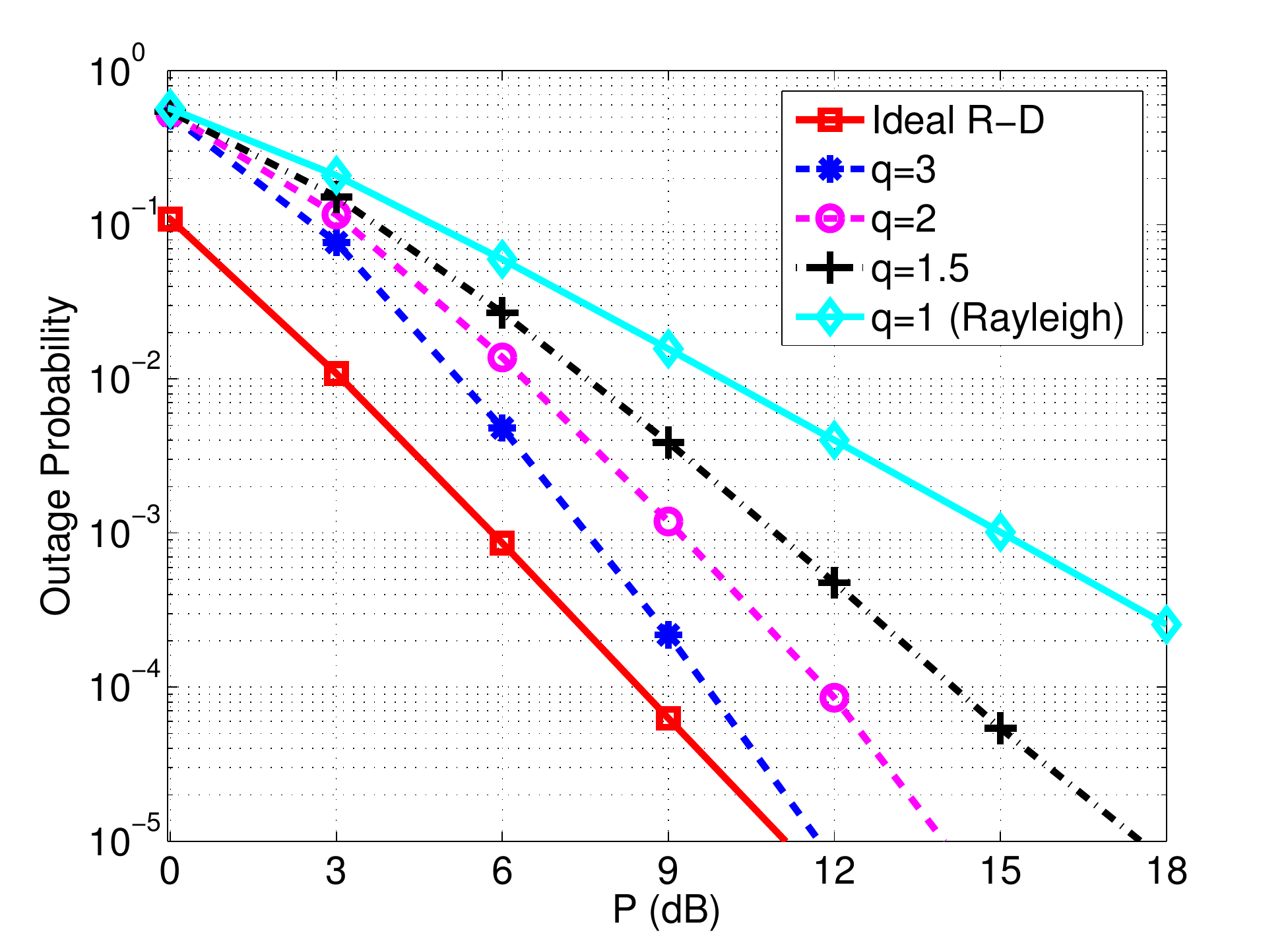}
\caption{\small Outage probability of the successive CMF method with D-R feedback, versus average SNR, for Nakagami($q$) R-D channels, ($R_{\text{t}}=2$, $L=4$, $M=2$,               $N=4$).}
\label{fig:SucCMF_WF}
\end{figure}

Fig.~\ref{fig:SucCMF_WF} considers the case of non-ideal R-D channels with Nakagami($q$) distribution, for different values of the channel parameter $q$. The corresponding outage probabilities of the successive CMF method with D-R feedback, versus average SNR, are shown. The ideal channel case is also plotted for comparison. The parameters $R_{\text{t}}=2$, $L=4$, $M=2$,               $N=4$ are assumed. As it is found from the figure, for $q \le 2$, as the fading severity decreases, i.e. $q$ increases, higher order of diversity is achieved. However, for $q>2$, a fixed diversity order, nearly $4$, is observed. This is due to the fact that the diversity order, from~(\ref{eq:d-WF}), is equal to  $\min \left( {Mq,\frac{MN}{2}} \right) $. Hence, for this figure, the diversity order is also limited by $\frac{MN}{2}=4$.

\begin{figure}[tb]
\renewcommand{\figurename}{Fig.}
\centering
\includegraphics[trim = 6mm 2mm 15mm 6mm, clip, width =\Wf \columnwidth]{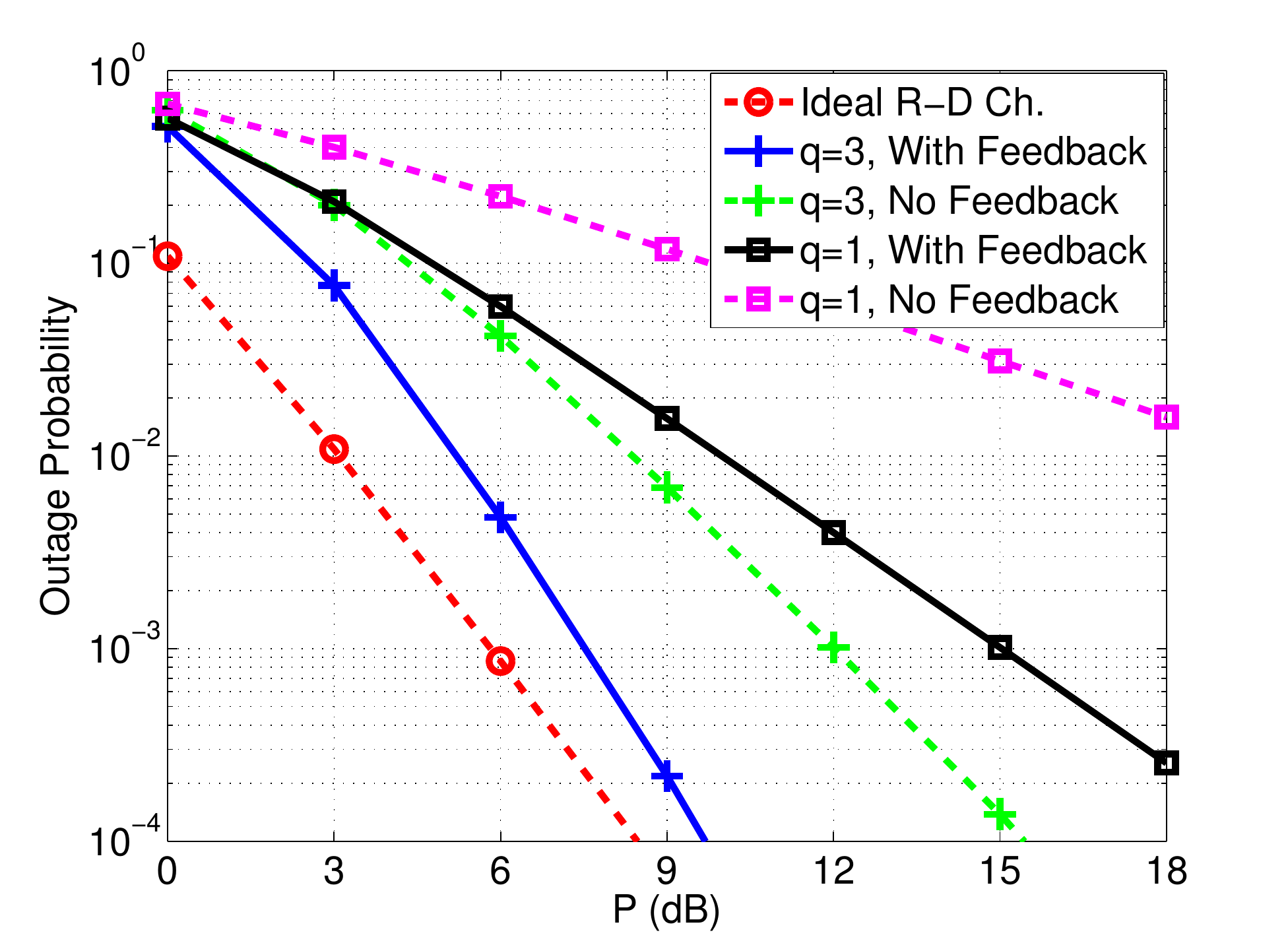}
\caption{\small Effect of using D-R feedback on outage probability of the successive CMF method, versus average SNR, for Nakagami($q$) R-D channels, ($R_{\text{t}}=2$, $L=4$, $M=2$, $N=4$).}
\label{fig:SucCMF_NF_WF}  
\end{figure}

In Fig.~\ref{fig:SucCMF_NF_WF}, the effect of using D-R feedback on the outage probability of the successive CMF method is shown for the case of non-ideal R-D channels. Ideal and Nakagami($q$) R-D channels with $q=1,3$, are considered. Parameters $R_{\text{t}}=2$, $L=4$, $M=2$, $N=4$ are selected. As it is observed, using D-R feedback for non-ideal R-D channels, improves the outage performance considerably, since it provides diversity gains. For instance, using D-R feedback for the Nakagami($q=1$) R-D channels, changes diversity order from $1$ to about $2$. This is in harmony with our theoretical diversity analysis in~(\ref{eq:d-NF}) and~(\ref{eq:d-WF}), that state diversity orders as $\min \left( {q,\frac{MN}{2}} \right) $ and $\min \left( {Mq,\frac{MN}{2}} \right) $ for the cases of no feedback and with feedback, respectively.

\begin{figure}[tb]
\renewcommand{\figurename}{Fig.}
\centering
\includegraphics[trim = 2mm 2mm 15mm 6mm, clip, width =\Wf \columnwidth]{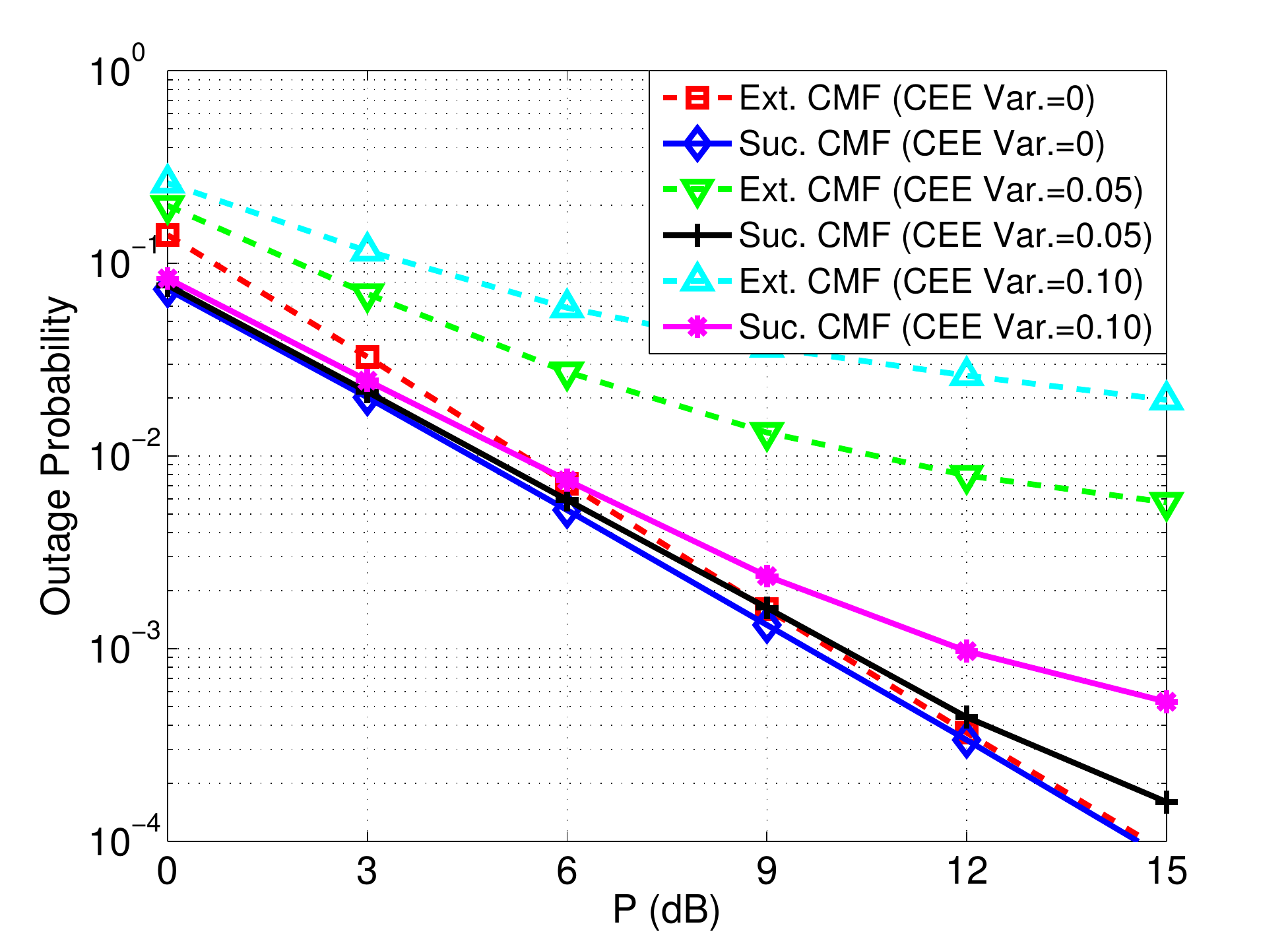}
\caption{\small Effect of channel estimation error on outage probability of the extended and successive CMF methods, versus average SNR, (ideal R-D channels, $R_{\text{t}}=1$, $L=4$, $M=1$, $N=4$).}
\label{fig:CEE}
\end{figure}

Fig.~\ref{fig:CEE} shows the effect of CEE on the outage performance of the extended and successive CMF methods. Parameters $R_{\text{t}}=1$, $L=4$, $M=1$, $N=4$ are selected. Ideal R-D channels are considered. Due to the time varying nature of wireless fading channels and non-ideal channel estimation methods, the CSI known at the relays, required for CMF methods, contains error. To study the effect of CEE, we model the estimated channel gains $h_{ln}^m, \forall m,n,l,$ as
\begin{equation} \label{eq:CEE}
\hat h_{ln}^m = \sqrt {1 - \sigma _e^2} h_{ln}^m + {\sigma _e}\varepsilon _{ln}^m,
\end{equation}
where $\varepsilon _{ln}^m$ is a real zero-mean Gaussian distribution with unit variance independent of true channel gain $h_{ln}^m$, and $\sigma _e^2$ denotes the CEE variance. As it is observed from Fig.~\ref{fig:CEE}, CEE deteriorates the performance of the both methods. However, successive CMF method is considerably more robust than the extended CMF method against the CEE.

To summarize, we have compared the discussed CMF methods in Table~\ref{tb:Compare}.

\section{Conclusion} \label{sec:concl}
In this paper, we proposed two novel CMF methods, namely, extended CMF and successive CMF methods, for multi-user multi-relay networks. Both methods exploit the new Sel-FW strategy that is applicable in the networks with limitations on signaling overheads and for arbitrary number of users/relays. Moreover, the proposed CMF methods require only local CSI at the relays.
The extended CMF and successive CMF methods exploit the Ext-CM and Suc-CM, respectively, as their computing scheme.
We formulated the Suc-CM scheme in a concise form and presented the explicit frameworks for selecting ECVs and decoding equations at relays. Furthermore, we suggested an approach to simplify the Suc-CM optimization problem to the regular form appeared in the original CMF method. 

Both extended and successive CMF methods solve the rank failure problems and outperform the original CMF method of~\cite{BN1} with a significant gap in outage probability. 
By theoretical analysis and computer simulations, we showed that both methods can achieve full diversity of the network, i.e. $\frac{MN}{2}$, provided that the R-D channels are stronger than a certain threshold, i.e. $q \ge \frac{MN}{2}$ for the case of no D-R feedback and $q \ge \frac{N}{2}$ for the case of using D-R feedback.
Our simulation results indicate that the successive CMF method provides SNR gains
 and considerable robustness against CEE, compared to the extended CMF method.

\appendices
\section{Proof of Theorem~\ref{th:IFLR}} \label{ap:IFLR}
The proof is based on the contradiction. The channel matrix $\mathbf{G}$ is given. Let the vectors in ${{\mathbf{A}}^*} = \left\{ {{{\mathbf{a}}_1}, \ldots ,{{\mathbf{a}}_L}} \right\}$ be the ECVs selected in turn by the Ext-CM optimization in~(\ref{eq:ak}). From~(\ref{eq:R_G_a}), their corresponding computation rates are $R({{\mathbf{a}}_k}) \triangleq R({\mathbf{G}},{{\mathbf{a}}_k}),\,k = 1, \ldots ,L$. From~(\ref{eq:ak}), these ECVs are linearly independent and we have $R({{\mathbf{a}}_1}) \geqslant  \ldots  \geqslant R({{\mathbf{a}}_L})$. Hence, the sum rate of Ext-CM scheme is  $R_{\text{sum}}^{\mathrm{(Ext)}} = L \times R({{\mathbf{a}}_L})$.

Suppose that the set ${\Omega ^*} = \left\{ {{{\mathbf{d}}_1}, \ldots ,{{\mathbf{d}}_L}} \right\}$ is the optimum solution of the IFLR optimization in~(\ref{eq:Rsum-IFLR}). From~(\ref{eq:Rsum-IFLR}), the $L$ ECVs in ${\Omega ^*} $ are linearly independent and span of ${\Omega ^*} $ has the rank of $L$, i.e. $\mathrm{rank}({\Omega ^*})=L$. Without loss of generality, assume that the ECVs in ${\Omega ^*}$ are sorted in descending order of their computation rates, i.e. we have $R({{\mathbf{d}}_1}) \geqslant  \ldots  \geqslant R({{\mathbf{d}}_L})$. Thus, the sum rate of IFLR scheme is $R_{\text{sum}}^{({\text{IFLR}})} = L \times R({{\mathbf{d}}_L})$. 

Now, assume that $R_{\text{sum}}^{({\text{IFLR}})}  >  R_{\text{sum}}^{\mathrm{(Ext)}}$. Hence, we can write
\begin{equation} \label{eq:temp05}
R({{\mathbf{d}}_1}) \geqslant  \ldots  \geqslant R({{\mathbf{d}}_L}) > R({{\mathbf{a}}_L}).
\end{equation} 

From~(\ref{eq:R_G_a}) and~(\ref{eq:ak}), ${{\mathbf{a}}_L}$ is given by
\begin{equation} \label{eq:ak04}
\begin{aligned}
{{\mathbf{a}}_L} &= \arg \max_{\substack{{\mathbf{a}} \in {\mathbb{Z}^L},{\mathbf{a}} \ne {\mathbf{0}}\\
{\mathbf{a}} \perp\!\!\!\perp \left\{ {{{\mathbf{a}}_1}, \ldots ,{{\mathbf{a}}_{L - 1}}} \right\}}}
 {R({\mathbf{G}},{\mathbf{a}})} \\ 
 &= \arg \max_{\substack{{\mathbf{a}} \in {\mathbb{Z}^L},{\mathbf{a}} \ne {\mathbf{0}}\\
{\mathbf{a}} \notin span\left(  \left\{ {{{\mathbf{a}}_1}, \ldots ,{{\mathbf{a}}_{L - 1}}} \right\} \right)}}
 {R({\mathbf{G}},{\mathbf{a}})} . 
\end{aligned}
\end{equation}

Note that ${{\mathbf{a}}_L}$ is not necessarily the unique solution of~(\ref{eq:ak04}), in the sense that there may be other ECVs that have a computation rate equal to   $R({{\mathbf{a}}_L})$. However,  from~(\ref{eq:ak04}),  for every ECV $\mathbf{b}$ such that ${\mathbf{b}} \notin span\left(  \left\{ {{{\mathbf{a}}_1}, \ldots ,{{\mathbf{a}}_{L - 1}}} \right\} \right)$ we have $R({\mathbf{b}}) \leq R({{\mathbf{a}}_L})$. In other words, for every ECV ${{\mathbf{c}}}$ with the computation rate $R({\mathbf{c}}) > R({{\mathbf{a}}_L})$, we have ${\mathbf{c}} \in span\left( {\left\{ {{{\mathbf{a}}_1}, \ldots ,{{\mathbf{a}}_{L - 1}}} \right\}} \right)$. Thus, from~(\ref{eq:temp05}), it is found that  ${\mathbf{d}_l} \in span\left( {\left\{ {{{\mathbf{a}}_1}, \ldots ,{{\mathbf{a}}_{L - 1}}} \right\}} \right)$, for $l=1,\ldots,L$. Hence, we can write ${\Omega ^*} \subseteq span\left( {\left\{ {{{\mathbf{a}}_1}, \ldots ,{{\mathbf{a}}_{L - 1}}} \right\}} \right)$.
As a result, we have  $\mathrm{rank}({\Omega ^*}) \leqslant L - 1$. This contradicts the assumption of $\mathrm{rank}({\Omega ^*}) = L$. Therefore, we have $R_{\text{sum}}^{({\text{IFLR}})}  =  R_{\text{sum}}^{\mathrm{(Ext)}}$.

\section{Proof of Theorem~\ref{th:b_opt}} \label{ap:bk_opt_proof}
Define the denominator of the computation rate in~(\ref{eq:R_G_b_a_sic}) as the function
\begin{equation}
f(\mathbf{b}_k,\pmb{\beta}_k)\triangleq \lVert\mathbf{b}_k\rVert ^2+\lVert\mathbf{G}\mathbf{b}_k+\mathbf{A}_{k-1}\pmb{\beta}_k-\mathbf{a}\rVert ^2 ,
\end{equation}
which can be expressed as
\begin{equation}
f(\pmb{\tau})= \lVert\pmb{\Sigma}\pmb{\tau}\rVert ^2+\lVert\pmb{\Gamma}\pmb{\tau}-\mathbf{a}\rVert ^2 ,
\end{equation}
where 
\begin{equation}
\begin{aligned}
\pmb{\tau}&\triangleq\begin{bmatrix}
\mathbf{b}_k\\
\pmb{\beta}_k
\end{bmatrix}_{(N+k-1)\times 1} ,\\
\pmb{\Gamma}&\triangleq\begin{bmatrix}
\mathbf{G}\mid
\mathbf{A}_{k-1}
\end{bmatrix}_{L\times (N+k-1)} ,\\
\pmb{\Sigma}&\triangleq\begin{bmatrix}
\mathbf{I}_N&\mathbf{0}\\
\mathbf{0}&\mathbf{0}
\end{bmatrix}_{(N+k-1)\times(N+k-1)} .
\end{aligned}
\end{equation}
Since $f(\pmb{\tau})$ can be rewritten as
a quadratic function in $\pmb{\tau}$, we can find  its minimum by setting its first derivative to zero, i.e.

\begin{equation}
\frac{\partial f}{\partial \pmb{\tau}}=2\pmb{\Sigma}^T\pmb{\Sigma}\pmb{\tau}+2\pmb{\Gamma}^T(\pmb{\Gamma}\pmb{\tau}-\mathbf{a})=0 ,
\end{equation}
that results in
\begin{equation}
\pmb{\tau}=(\pmb{\Sigma}+\pmb{\Gamma}^T\pmb{\Gamma})^{-1}\pmb{\Gamma}^T\mathbf{a} .
\end{equation}
Finally, using the  block matrix inversion relation~\cite[Sec. 3.7]{Meyer}, to find the inverse  of the matrix
\begin{equation}
\pmb{\Sigma}+\pmb{\Gamma}^T\pmb{\Gamma}=\begin{bmatrix}
\mathbf{I}_N+\mathbf{G}^T\mathbf{G}&\mathbf{G}^T\mathbf{A}_{k-1}\\
\mathbf{A}_{k-1}^T\mathbf{G}&\mathbf{A}_{k-1}^T\mathbf{A}_{k-1}
\end{bmatrix} ,
\end{equation}
and some manipulations, we get the desired result of the theorem.

\section{Lattice basics and preliminary definitions required for Section~\ref{subsec:Suc-opt}} \label{ap:definitions}

\begin{definition} \label{def:lattice}
A \textit{lattice} $\Lambda$ is a subgroup of $\mathbb{R}^L$ that for all vectors $\mathbf{x}, \mathbf{y} \in \Lambda$ we have $\mathbf{x}\pm \mathbf{y} \in \Lambda$~\cite[Ch. 1]{LatticeBook}. If all lattice points can be expressed as integer linear combinations of a set of vectors $\left\{\mathbf{x}_1,\ldots,\mathbf{x}_n \right\}$ in $\mathbb{R}^L$, i.e.
\begin{equation} \label{eq:lattice}
\Lambda  = \left\{ {\sum_{i = 1}^n {{a_i}{{\mathbf{x}}_i}} \left| {{a_1}, \ldots ,{a_n} \in \mathbb{Z}} \right.} \right\},
\end{equation}
the matrix ${\mathbf{X}} = \left[ {{{\mathbf{x}}_1}, \ldots ,{{\mathbf{x}}_n}} \right]$ and $K=\mathrm{rank}(\mathbf{X})\leq L$ are called generator matrix and the rank of lattice $\Lambda$, respectively. If $K=n$, i.e. the columns of $\mathbf{X}$ are linearly independent, the matrix $\mathbf{X}$ is called an standard generator matrix of lattice $\Lambda$; in this case every lattice point ${\mathbf{v}} \in \Lambda $ has a unique representation as ${\mathbf{v}} = {\mathbf{Xw}},{\mathbf{w}} \in {\mathbb{Z}^K}$~\cite[Ch. 1]{Murray}.
\end{definition}
\begin{definition}\label{def:unimodular}
A \textit{unimodular column  operation} on a matrix is one of the following elementary column  operations~\cite[Ch.1]{Murray}: multiply any column by $-1$, interchange any two columns, and add an integer multiple of a column to any other column.\\
The transformation matrix of a series of unimodular column operations on a matrix is given by a unimodular matrix $\mathbf{U}$. An $n \times n$ matrix with integer entries and determinant $\pm 1$ is called a \textit{unimodular matrix}.
\end{definition}
\begin{remark} \label{Re:genMat}
The generator matrix and the standard generator matrix of a lattice are not unique. Specifically, if $\mathbf{X}_1$ is a generator matrix for lattice $\Lambda$, then the matrix $\mathbf{X}_2=\mathbf{X}_1 \mathbf{U}$ is also a generator matrix for $\Lambda$, where $\mathbf{U}$ is a unimodular matrix~\cite[Ch. 1]{Murray}. Note that $\mathbf{X}_2$ can be generated by applying unimodular column operations on matrix $\mathbf{X}_1$.
\end{remark}
\begin{definition} \label{def:complement}
For a set of vectors $\left\{ \mathbf{x}_1,\ldots,\mathbf{x}_n \right\}$ in $\mathbb{R}^L$, $V=\mathrm{span}\left( \mathbf{x}_1,\ldots,\mathbf{x}_n \right)$ is the subspace spanned by linear combinations of these vectors. The \textit{orthogonal complement} $V^\perp$ , which is a subspace of $\mathbb{R}^L$, is defined as the set of all vectors in $\mathbb{R}^L$ that are orthogonal to every vector in~$V$~\cite[Ch. 5]{Meyer}.
\end{definition}

\section{Proof of Theorem~\ref{th:Opt_sic}} \label{ap:Opt_sic_proof}
By replacing the search set of optimization in~(\ref{eq:min_sic}) with~(\ref{eq:S2}), we have
\begin{equation} \label{eq:min_sic2}
\min_{\substack{{\mathbf{d}} = {{{\mathbf{\tilde P}}}_{k - 1}}{\mathbf{w}},{\mathbf{w}} \in {\mathbb{Z}^{L - k + 1}},  \\
{\mathbf{w}} \ne {\mathbf{0}}}} {{{\mathbf{d}}^{{T}}}{\mathbf{C}_k \mathbf{d}}}.
\end{equation} 
Hence, the optimum solution of~(\ref{eq:min_sic2}) is given by
\begin{equation} \label{eq:wk_sic}
\begin{aligned}
{{\mathbf{w}}_k} &= \arg \min_{\substack{{\mathbf{w}} \in {\mathbb{Z}^{L - k + 1}},  \\
{\mathbf{w}} \ne {\mathbf{0}}}} {{{\mathbf{w}}^T}{\mathbf{\tilde P}}_{k - 1}^T{{\mathbf{C}}_k}{{\mathbf{\tilde P}}_{k - 1}}{\mathbf{w}}},\\
 &= \arg \min_{\substack{{\mathbf{w}} \in {\mathbb{Z}^{L - k + 1}},  \\
{\mathbf{w}} \ne {\mathbf{0}}}} {{{\mathbf{w}}^T}{{\mathbf{\tilde Q}}_k}{\mathbf{w}}}.
\end{aligned}
\end{equation} 

Since ${{\mathbf{\tilde P}}_{k - 1}}{{\mathbf{w}}_k}$ is a vector in the search set $S$, from~(\ref{eq:S}), we have
\begin{equation}
{{{\mathbf{\tilde P}}}_{k - 1}}{{\mathbf{w}}_k} = {{\mathbf{F}}_{k - 1}}{{\mathbf{a}}_k},
\end{equation}
and from~(\ref{eq:Ptilde_k-1}), we can write
\begin{equation} \label{eq:FUw}
{{\mathbf{F}}_{k - 1}}{{\mathbf{U}}_{k - 1}}{{\mathbf{w}}_k} = {{\mathbf{F}}_{k - 1}}{{\mathbf{a}}_k}.
\end{equation}
From Remark~\ref{Re:F_k-1}, (\ref{eq:FUw}) indicates that two vectors ${{\mathbf{U}}_{k - 1}}{{\mathbf{w}}_k}$ and $\mathbf{a}_k$ have the same projection onto the orthogonal complement of $\mathrm{span}\left( \mathbf{a}_1,\ldots,\mathbf{a}_{k-1} \right)$. Thus, we have ${{\mathbf{a}}_k} = {{\mathbf{U}}_{k - 1}}{{\mathbf{w}}_k} + {{\mathbf{a}}_0}$ where ${{\mathbf{a}}_0} \in {\mathbb{Z}^L}$ can be any of the integer vectors in $\mathrm{span}\left( \mathbf{a}_1,\ldots,\mathbf{a}_{k-1} \right)$. Note that from~(\ref{eq:R_G_a_sic}), it can be shown that any choices of ${\mathbf{a}}_k$ give the same rate. Hence, we set ${{\mathbf{a}}_0}=\mathbf{0}$ to get the desired result.

Now we prove that ${{\mathbf{\tilde Q}}_k}$ is a positive definite matrix. As stated above, the columns of ${\mathbf{\tilde P}}_{k - 1}$ are linearly independent.
Hence for any ${\mathbf{x}} \ne {\mathbf{0}}$ in $\mathbb{R}^{L-k+1}$ we have $\mathbf{y} \triangleq {\mathbf{\tilde P}}_{k - 1} \mathbf{x}  \ne {\mathbf{0}}$. 
Therefore, we can write
\begin{equation}
\begin{aligned}
{{\mathbf{x}}^T}{{\mathbf{\tilde Q}}_k}{\mathbf{x}} &= {{\mathbf{x}}^T}{{\mathbf{\tilde P}}_{k - 1}}^T{{\mathbf{C}}_k}{{\mathbf{\tilde P}}_{k - 1}}{\mathbf{x}} \\
&= {{\mathbf{y}}^T}{{\mathbf{C}}_k}{\mathbf{y}} > 0,
\end{aligned}
\end{equation}
where the last inequality follows from the fact that ${{\mathbf{C}}_k} = {\left( {{\mathbf{I}_L} + {{{\mathbf{\tilde G}}}_k}{\mathbf{\tilde G}}_k^T} \right)^{ - 1}}$ is a positive definite matrix (Lemma~\ref{lem:Q_k}).

\section{Proof of Theorem~\ref{th:d-Ideal}} \label{ap:d-Ideal}
The relays in network exploit the Ext-CM or Suc-CM scheme. Hence, from Lemma~\ref{lem:d-SucCM} and definition of diversity order in~(\ref{eq:d-power}), the outage probability of the relay $m, 1 \leq m \leq M$, for $\gamma \to \infty$,  can be written as
\begin{equation}
P_{\text{relay},m}^{\text{out}} \doteq \frac{{{{\alpha '}_m}}}{{{\gamma ^{\frac{N}{2}}}}}\,\,\,m=1, \ldots,M,
\end{equation}
where $\gamma$ is the average SNR, and ${{\alpha '}_m}, m=1,\ldots,M,$ are constants (independent of $\gamma$). 
Thus, from~(\ref{eq:Pout_sys_Ideal}), the outage probability of the system is given by
\begin{equation} \label{eq:Pout-sys-Ideal01}
P_{\text{sys,Ideal}}^{\text{out},M} \doteq \prod\nolimits_{m = 1}^M {\frac{{{{\alpha '}_m}}}{{{\gamma ^{\frac{N}{2}}}}}}  \doteq \frac{{{\alpha _0}}}{{{\gamma ^{\frac{{MN}}{2}}}}},
\end{equation}
where $\alpha _0$ is a constant. From~(\ref{eq:Pout-sys-Ideal01}) and definition of diversity order, we get ${d_{\text{sys,Ideal}}} = \frac{{MN}}{2}$.

\section{Proof of Theorem~\ref{th:d-NF}} \label{ap:d-NF}
From~(\ref{Pout-fm}), we have
\begin{equation}
\mathop {\lim }\limits_{\gamma  \to \infty } \frac{{P_{\text{R-D}}^{\text{out}}}}{\gamma } = \mathop {\lim }\limits_{\gamma  \to \infty } \frac{1}{\gamma }{F_{{\chi ^2}}}\left( {2q,\frac{{{2^{\frac{{2{R_{\text{t}}}}}{L}}} - 1}}{\gamma }} \right) = q .
\end{equation}
Hence, from the definition of diversity, in~(\ref{eq:d-power}), we can write
\begin{equation}\label{eq:Pout_R-D02}
P_{\text{R-D}}^{\text{out}} \doteq \frac{{{\alpha _3}}}{{{\gamma ^q}}},
\end{equation}
at high SNRs, where $\alpha _3$ is a constant (independent of SNR $\gamma$). Thus, from~(\ref{eq:Pout-sys-NF}) and~(\ref{eq:Pout-sys-Ideal01}), the system outage probability is given by
\begin{equation}
\begin{aligned}
  P_{\text{sys,NF}}^{\text{out}} &\doteq \frac{{{\alpha _3}}}{{{\gamma ^q}}} + \left( {1 - \frac{{{\alpha _3}}}{{{\gamma ^q}}}} \right)\frac{{{\alpha _0}}}{{{\gamma ^{\frac{{MN}}{2}}}}}  \\
   &\doteq \frac{{{\alpha _4}}}{{{\gamma ^{\min \left( {q,\frac{{MN}}{2}} \right)}}}}  ,
\end{aligned}
\end{equation}
where $\alpha _4$ is a constant. From the definition of diversity order, this yields the desired result.

\section{Proof of Theorem~\ref{th:d-WF}} \label{ap:d-WF}
From definition of diversity order in~(\ref{eq:d-power}) and by substituting~(\ref{eq:Pout-sys-Ideal01}) and~(\ref{eq:Pout_R-D02}) in~(\ref{eq:Pout-WF1}), we have
\begin{equation} \label{eq:temp03}
\begin{aligned}
P_{\text{sys,WF}}^{\text{out}} &\doteq \sum\limits_{m = 0}^M {{M \choose m}{{\left( {1 - \frac{{{\alpha _3}}}{{{\gamma ^q}}}} \right)}^m}{{\left( {\frac{{{\alpha _3}}}{{{\gamma ^q}}}} \right)}^{M - m}}\frac{{{{\alpha ''}_m}}}{{{\gamma ^{\frac{{mN}}{2}}}}}} \\
&\doteq \sum\limits_{m = 0}^M {\frac{{{{\alpha '''}_m}}}{{{\gamma ^{q\left( {M - m} \right) + \frac{{mN}}{2}}}}}} \\
&\doteq \frac{{{\alpha _5}}}{{{\gamma ^{{d_{\text{sys,WF}}}}}}},
\end{aligned}
\end{equation}
where $\alpha _5$, ${\alpha''} _m, m=1,\ldots,M$, and ${\alpha'''} _m,m=1,\ldots,M$, are some constants, and ${{d_{\text{sys,WF}}}}$ is the diversity order. From~(\ref{eq:temp03}), ${{d_{\text{sys,WF}}}}$ is calculated as
\begin{equation}
\begin{aligned}
  {d_{\text{sys,WF}}} &= \mathop {\min }\limits_{0 \leqslant m \leqslant M} q\left( {M - m} \right) + \frac{{mN}}{2} \\
  &= \mathop {\min }\limits_{0 \leqslant m \leqslant M} \frac{{2qM + (2q - N)m}}{2}  \\
  & = \min \left( {qM,\frac{{MN}}{2}} \right) \\
  &= M \times \min \left( {q,\frac{N}{2}} \right) . 
\end{aligned}
\end{equation}


\section*{Acknowledgment}

This work has been supported in part by the VR research link project ``Green Communication via Multi-relaying.''

\ifCLASSOPTIONcaptionsoff
  \newpage
\fi

\end{document}